\documentclass[%
 prl,
 amsmath,amssymb,
 superscriptaddress
]{revtex4-2}
\usepackage{lipsum}
\usepackage{balance}
\usepackage{graphicx}
\newcommand{\byline}[1]{{\em #1 \\}\twocolumngrid}
\newcommand{\ctitle}[1]{\onecolumngrid\clearpage\noindent{\Large #1 \\}}

\usepackage[utf8]{inputenc}
\usepackage[colorlinks=true,linkcolor=blue]{hyperref}
\newcommand{\alfa}{\alpha}
\newcommand{\xb}{x_{\mbox{\tiny\!$B$}}}
\newcommand{\xL}{x_{\mbox{\tiny $L$}}}
\newcommand{\zL}{z_{\mbox{\tiny $L$}}}
\newcommand{\xh}{x_h}
\newcommand{\diff}[1]{\mathrm{d}#1}
\newcommand{\fref}[1]{Fig.~\ref{fig:#1}}

\begin{document}
\title{CFNS Ad-Hoc meeting on Radiative Corrections Whitepaper}

\author{Andrei Afanasev}
\affiliation{The George Washington University, Washington, DC}
\author{Jaseer Ahmed}
\affiliation{University of Manitoba, Winnipeg, MB Canada}
\author{Igor Akushevich}
\affiliation{Physics Department, Duke University, Durham, NC}
\affiliation{Jefferson Lab., Newport News, VA} 
\author{Jan C.~Bernauer}
\affiliation{Stony Brook University, Stony Brook, NY}
\affiliation{Riken BNL Research Center, Upton, NY}
\author{Peter G.~Blunden}
\affiliation{University of Manitoba, Winnipeg, MB Canada}
\author{Andrea Bressan}
\affiliation{University of Trieste, Dept. of Physics, Trieste, Italy}
\affiliation{Trieste Section of INFN, Trieste, Italy}
\author{Duane Byer}
\affiliation{Duke University, Durham, NC}
\author{Ethan Cline}
\affiliation{Stony Brook University, Stony Brook, NY}
\author{Markus Diefenthaler}
\affiliation{Jefferson Lab., Newport News, VA}
\author{Jan M.~Friedrich}
\affiliation{Technische Universität München, Physik Dept., Garching, Germany}
\author{Haiyan Gao}
\affiliation{Duke University, Durham, NC}
\author{Alexandr Ilyichev}
\affiliation{Institute for Nuclear Problems, Belarusian State University, Minsk, Belarus}
\author {Ulrich D.~Jentschura}
\affiliation{Missouri University of Science 
and Technology, Rolla, MO}
\author{Vladimir Khachatryan}
\affiliation{Duke University, Durham, NC}
\author{Lin Li}
\affiliation{University of South Carolina, Columbia, SC}
\author{Wally Melnitchouk}
\affiliation{Jefferson Lab, Newport News, VA}
\author{Richard Milner}
\affiliation{Massachusetts Institute of Technology, Cambridge, MA}
\author{Fred Myhrer} 
\affiliation{University of South Carolina, Columbia, SC} 
\author{Chao Peng}
\affiliation{Argonne National Laboratory, Lemont, IL}
\author{Jianwei Qiu}
\affiliation{Jefferson Lab, Newport News, VA}
\author{Udit Raha}
\affiliation{Indian Institute of Technology Guwahati, Guwahati, Assam, India.}
\author{Axel Schmidt}
\affiliation{The George Washington University, Washington, DC}
\author{Vanamali C.~Shastry}
\affiliation{Regional Institute of Education Mysuru, Mysore, India}
\author{Hubert Spiesberger}
\affiliation{PRISMA+ Cluster of Excellence, Institut fur Physik, Johannes Gutenberg Universitat Mainz, Mainz, Germany}
\author{Stan Srednyak}
\affiliation{Duke University, Durham, NC}
\author{Steffen Strauch}
\affiliation{University of South Carolina, Columbia, SC}
\author{Pulak Talukdar}
\affiliation{Indian Institute of Technology Guwahati, Guwahati, Assam, India.}

\author{Weizhi Xiong}
\affiliation{Syracuse University, Syracuse, NY}

\maketitle

\ctitle{Foreword}
\byline{Jan C. Bernauer, Jan Friedrich, Haiyan Gao, Richard Milner}
Current precision scattering experiments and even more so many experiments planed for the Electron Ion Collider will be limited by systematics.  From the theory side, a fundamental source of systematic uncertainty is the correct treatment of radiative effects. To gauge the current state of technique and knowledge, help the cross-pollination between different direction of nuclear physics, and to give input to the yellow report process, the community met in an ad-hoc workshop hosted by the Center for Frontiers in Nuclear Science, Stony Brook University. This whitepaper is a collection of contributions to this workshop. 

Radiative effects are classically treated by introducing correction factors to map the measured quantity back to the pure, radiation-effects-free base quantity. For example, a proton form factor experiment actually measures a small energy range of inelastic scattering elements, i.e., those where only a small part of the energy is lost in photons compared to an elastic event. This rate is then mapped back to the elastic rate via correction factors. As can be seen here, this makes the correction not a purely theoretical quantity, as its size depends on the experimental setup, in this case, what energy range is accepted as elastic.  To achieve the highest precision, these effective cuts are often better handled not by applying a correction-factor post-hoc, but by integrating the radiative correct cross section into a Monte Carlo simulation. As such, radiative corrections are really at the interface between experiment and theory, and the problem must be attacked from both directions.

At the same time, experiment design can be affected by considerations of radiative corrections, a detail which is often overlooked. On one hand, choosing suitable detector technologies and experiment kinematics can minimize the uncertainty from radiative corrections, for example due to accurate knowledge of the acceptances relevant for the corrections. On the other hand, added detectors for radiated photons, for example, could reduce the contribution to the measured cross section or allow a calibration of the correction code. Accelerator capabilities like polarization or the change of the particle charge can also be used to measure, or eliminate, radiative corrections.

It is our impression that the field is making good progress on the theory front, with new codes for higher-order corrections becoming available, for example, for electron-proton scattering. The implementation of radiative corrections for jet physics seems also on a good pass, but is maybe somewhat further behind. In general, the path to EIC seems without major obstacles, but we miss a more broad discussion of detector design with regard to radiative corrections. Radiative corrections are often only an afterthought, and we hope that the EIC project will continue to give this important topic enough attention.

\ctitle{The current state of affairs}
\byline{Jan Friedrich}

Practically all scattering reactions of interest in the context of the standard model of particle physics involve the acceleration of electric charge, and modifications due to coupling to the electromagnetic field need to be taken into account in the interpretation of such scattering data.

Already early examinations of quantum electrodynamics have revealed the formal problems dubbed the ``infrared catastrophe" \cite{PhysRev.52.54} for the diverging emission probability for soft photons. The link to a part of the "ultraviolet" divergences was further worked out and lead to the successful first-order QED radiative correction scheme as given for electron scattering by Mo and Tsai \cite{Mo:1968cg}. For the higher orders, the cancellation of the formally arising divergences was shown, however ``in accordance with the usual spirit of perturbation theory," higher-order finite contributions are ignored "in spite of our complete ignorance of their convergence properties" \cite{Yennie:1961ad}. The related exponentiation procedure implies the respective uncertainty, unless its validity is proven for finite-energy photons order by order, as is also admitted in modern formulations as \cite{Passarino_2001}, that the "next-to-leading terms ... require instead a comparison with the explicit calculation of the cross-section up to the two-loop level". A complete evaluation of the two-loop level for elastic electron scattering is still not employed beyond the exponentiated form of the one-loop level.

The cross-section for multiple-photon emission, in principle calculable from QED to any order, determines the shape of the radiative tail associated with the elastic process, and especially the region close to the non-radiative elastic kinematics. Experimentally, these effects come together with those of external bremsstrahlung and energy loss of the electrons due to atomic ionisation in the target material. A precise treatment is only possible via Monte Carlo simulations, if credible models for all the contributions are available. For external bremsstrahlung and ionisation energy loss, the model precision is limited by the experimental uncertainties on their empiric parameters \cite{friedrich2000messung}, and by approximations made for computational efficiency. Regarding the internal photon emission processes, modern generators such as ESEPP \cite{Gramolin:2014pva} realize only first-order photon emission, based on a photon energy cutoff. This energy cutoff is often argued together with the detector resolution---indeed it does not make much sense to aim at describing features of the peak shape far beyond what is experimentally observable---however the mentioned issues with the exponentiation procedure come on top, and cannot be consistently included in such event generators.

Interesting input can be expected from experiment, if in a well-determined process such as elastic lepton-proton scattering additional photons are observed. The used generators for including radiation effects can then be tested by comparing the observed angular distributions with the shapes expected from QED. Since the effects depend also on the lepton mass, it is foreseen in the Proton Radius Measurement with a 100~GeV muon beam by COMPASS++/AMBER \cite{Adams:2676885} at the CERN SPS M2 beam line, where the bremsstrahlung photons are strongly forward boosted.

Regarding corrections for higher-order contributions due to virtual photons, a similar control is possible by comparing measurements with leptons of positive and negative charge, as continued with the OLYMPUS experiment \cite{Henderson:2016dea} at DESY.

Key to the implementation of radiative corrections is the consistent split into the theory calculation to determine the "remainder effect" of the regularization procedure for the appearing singularities, and the part which is influenced by the effects of the experimental technique and the cuts applied in the analysis. This becomes a strongly increasing challenge if higher orders and multiple photon emission are to be incorporated. For all the experimental approaches discussed in this white paper, there is a common interest to implement higher-order QED effects in a form such that their analytic parts are well-defined, together with appropriate Monte Carlo methods for the simulation to correct the respective data.

Those efforts are of particular interest in the context of the upcoming Electron-Ion Collider \cite{accardi2014electron} where large effects must be expected due to internal radiation effects accompanying the processes of interest.

\newcommand\be{\begin{equation}}
\newcommand\ee{\end{equation}}
\newcommand\bea{\begin{eqnarray}}
\newcommand\eea{\end{eqnarray}}

\ctitle{Radiative corrections for precision $e$-$p$ scattering}
\byline{Jaseer Ahmed, Peter Blunden, Wally Melnitchouk}

For many decades the proton's electric
($G_E(Q^2)$) and magnetic ($G_M(Q^2)$) elastic form factors have been measured
in unpolarized scattering experiments using the Rosenbluth
longitudinal-transverse (LT) separation technique~\cite{walker1994,
Andivahis1994, qattan2005}. These experiments found that the ratio $\mu_p\,
G_E/G_M$, where $\mu_p$ is the proton's magnetic moment, is consistent with 1
over a large range of the four-momentum transfer squared, $Q^2$, up to
8.83~GeV$^2$. More recently, measurements of the electric to magnetic form
factor ratio with significantly reduced uncertainties were performed at
Jefferson Lab using the polarization transfer (PT) technique~\cite{jones2000g,
gayou2002,punjabi2005, puckett2010, puckett2012}. These experiments found a
linear fall-off of the ratio $\mu_p\, G_E/G_M$ from 1 with increasing $Q^2$ in
the range up to 8.5~GeV$^2$.

Analysis of the LT-separated 
electron scattering data has traditionally been
performed within the one-photon exchange (OPE) approximation.  The electric to
magnetic form factor ratio discrepancy motivated studies of hadron
structure-dependent two-photon exchange (TPE) radiative corrections, and it was
generally believed that the problem would be resolved with the inclusion of
these effects~\cite{blunden2003, guichon2003}.

The use of hadronic degrees of freedom can be considered as a reasonable
approximation for low to moderate values of $Q^2 \lesssim 5$~GeV$^2$, where
hadrons are expected to retain their identity. On-shell form
factors are used explicitly to calculate the imaginary part of the TPE amplitude
from unitarity, with the real part then obtained from a dispersion integral.
In our work~\cite{Ahmed2020} we follow the dispersive approach for resonant
intermediate states developed in Ref.~\cite{blunden2017}. 
We account for all four and three-star spin-$1/2^\pm$ and spin-$3/2^\pm$ resonances
with mass below 1.8~GeV from the Particle Data Group~\cite{pdg2018},
which include the six isospin-1/2 states
    $N(1440)$,
    $N(1520)$,
    $N(1535)$, 
    $N(1650)$, 
    $N(1710)$ and
    $N(1720)$,
and the three isospin-3/2 states
    $\Delta(1232)$,
    $\Delta(1620)$ and
    $\Delta(1700)$.
We allow for a Breit-Wigner shape
with a nonzero width for each individual resonance, with either a fixed width or
a dynamical width that depends on the final state hadron mass.
In our numerical calculations, for the resonance electrocouplings at the
hadronic vertices we use the most recent helicity amplitudes extracted from the
analysis of CLAS electroproduction data~\cite{mokeev2012, HillerBlin:2019hhz},
except for the 
$\Delta(1232)$ resonance, for which we use the fit by
Aznauryan and Burkert~\cite{blunden2017,aznauryan2012}. For the elastic
intermediate state contribution, we use
the parametrizations from Kelly~\cite{kelly2004}, which have poles only in the
timelike region of $Q^2$.

\begin{figure}[ht]
\includegraphics[width=0.45\textwidth]{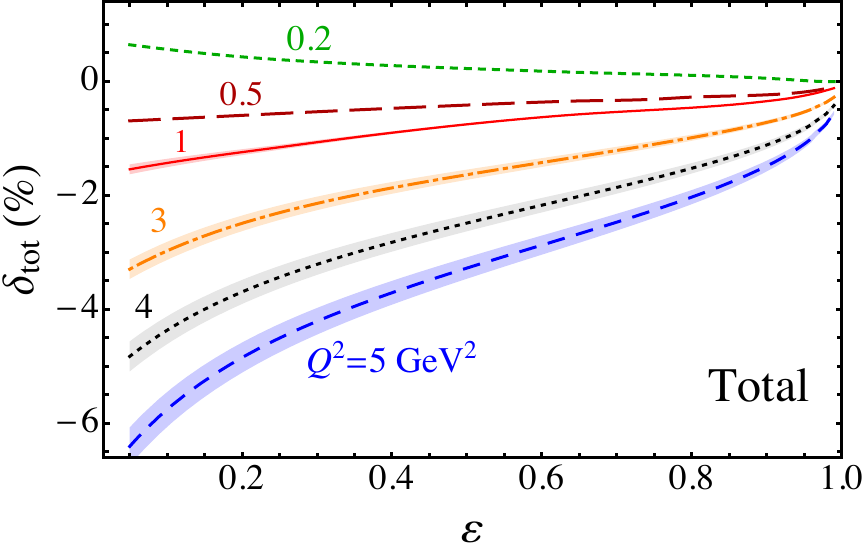}
\caption{The total TPE correction $\delta$ (in \%) versus  $\varepsilon$ for
nucleon plus all spin-parity $1/2^\pm$ and $3/2^\pm$ resonances at
    $Q^2 = 0.2$~GeV$^2$ (green dashed line),
    0.5~GeV$^2$ (dark red long-dashed),
    1~GeV$^2$ (red solid),
    3~GeV$^2$ (orange dot-dashed),
    4~GeV$^2$ (black dotted), and
    5~GeV$^2$ (blue dashed). The shaded bands correspond to
    the uncertainty propagated from the input electrocouplings.}
\label{fig.Dsig}
\end{figure}

The TPE correction $\delta$ is defined via
\be
\frac{d\sigma}{d\Omega}=\left(\frac{d\sigma}{d\Omega}\right)_{\rm Born}(1+\delta).
\ee
The combined effect on $\delta$ from the nucleon plus all the spin-parity
$1/2^\pm$ and $3/2^\pm$ resonances is illustrated in Fig.~\ref{fig.Dsig} as a
function of virtual photon polarization, $\varepsilon$, for a range of fixed
$Q^2$ values between 0.2 and 5~GeV$^2$. 
At low $Q^2 \lesssim 3$~GeV$^2$, the net excited state resonance contribution
is found to be small, and the total correction is dominated by the nucleon elastic
intermediate state.
The net effect of the higher mass resonances is to increase the magnitude of the
TPE correction at $Q^2 \gtrsim 3$~GeV$^2$, due primarily to the growth of the
(negative) odd-parity 
$N(1520)$ and $N(1535)$ resonances, which
overcompensates the (positive) contributions from the 
$\Delta(1232)$.
At the highest $Q^2=5$~GeV$^2$ value, the total TPE
correction $\delta_{\rm tot}$ reaches $\approx 6\%$-$7\%$ at low $\varepsilon$.

An estimate of the theoretical uncertainties on the TPE contributions can be made
by propagating the uncertainties on the fitted values of the transition
electrocouplings~\cite{HillerBlin:2019hhz}, which are dominated by the
$\Delta(1232)$ and $N(1520)$ intermediate states.
At low~$Q^2$, $Q^2 \lesssim 0.5$~GeV$^2$, the uncertainties are insignificant,
but become more visible at higher $Q^2$ values, as illustrated by the shaded bands
in Fig.~\ref{fig.Dsig}, for $Q^2 = 1$-$5~ \rm{GeV}^2$.

Perhaps the most direct consequence of TPE is the deviation
from unity of the ratio of $e^+ p$ to $e^- p$ elastic scattering cross sections,
which behaves like $R_{2\gamma} = \sigma(e^+ p)/\sigma(e^- p)
\approx\ 1 - 2\, \delta$, and is a direct
measure of the TPE correction $\delta$. Recent experiments at
Jefferson Lab~\cite{CLASTPE2017}, Novosibirsk~\cite{VEPP2015} and
DESY~\cite{OLYMPUS2017} have attempted more precise determinations of
$R_{2\gamma}$ over a larger range of $Q^2$ and $\varepsilon$ values than
previously available.

\begin{figure}[t]
\includegraphics[width=0.45\textwidth]{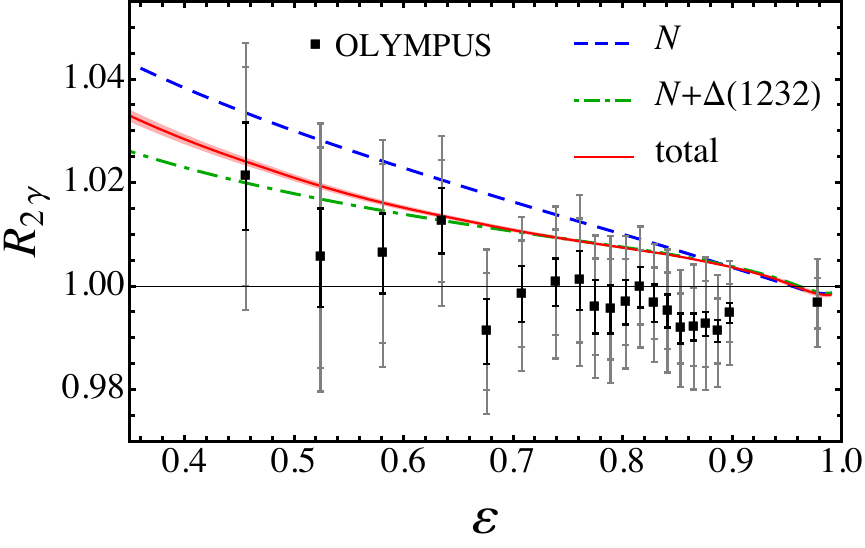}
\caption{Ratio $R_{2\gamma}$ versus
$\varepsilon$ from the Ref.~\cite{OLYMPUS2017}, compared with the nucleon only (blue dashed line), sum of nucleon
and $\Delta(1232)$ (green dot-dashed line), and sum of all intermediate state
contributions (red solid line and band). The experimental statistical and systematic
uncertainties are indicated by the (black) inner and (gray) outer error bars,
respectively.}
\label{fig.Olympus}
\end{figure}

The OLYMPUS experiment at DESY~\cite{OLYMPUS2017} measured the ratio
$R_{2\gamma}$ over a range of $\varepsilon$ from $\approx 0.46$ to 0.9 at an
electron energy $E \approx 2$~GeV, with $Q^2$ ranging up to $\approx 2$~GeV$^2$.
The results, illustrated in Fig.~\ref{fig.Olympus}, indicate an enhancement of
the ratio at $\varepsilon \lesssim 0.6$ and a dip below unity at $\varepsilon
\gtrsim 0.7$, although still compatible with no deviation from 1 within the
combined statistical and systematic uncertainties. The suppression of the ratio
at large $\varepsilon$ is in tension with other measurements, but the effect is consistent within the errors~\cite{blunden2017}. Inclusion of the
$\Delta(1232)$ intermediate state reduces the effect of the nucleon elastic
contribution away from the forward scattering region, but the effect of the
higher mass resonances is very small for all $\varepsilon$ shown. The overall
agreement between the TPE calculation and the OLYMPUS data is reasonable within
the experimental uncertainties, although there is no indication in our model for
a decrease of the ratio below unity at large~$\varepsilon$.

The most widely discussed consequence of TPE 
over the last two decades is the ratio $\mu_p\, G_E/G_M$ extracted using the LT separation
method~\cite{blunden2003}. The TPE correction induces an additional
shift in the $\varepsilon$ dependence of the reduced cross section, which affects
the extraction of both $G_E$ and $G_M$. To extract $G_E$ and $G_M$ it
is appropriate to correct the data for TPE contributions at the same level
as other radiative corrections in order to obtain the genuine Born contribution.

\begin{figure}[b]
\includegraphics[width=0.42\textwidth]{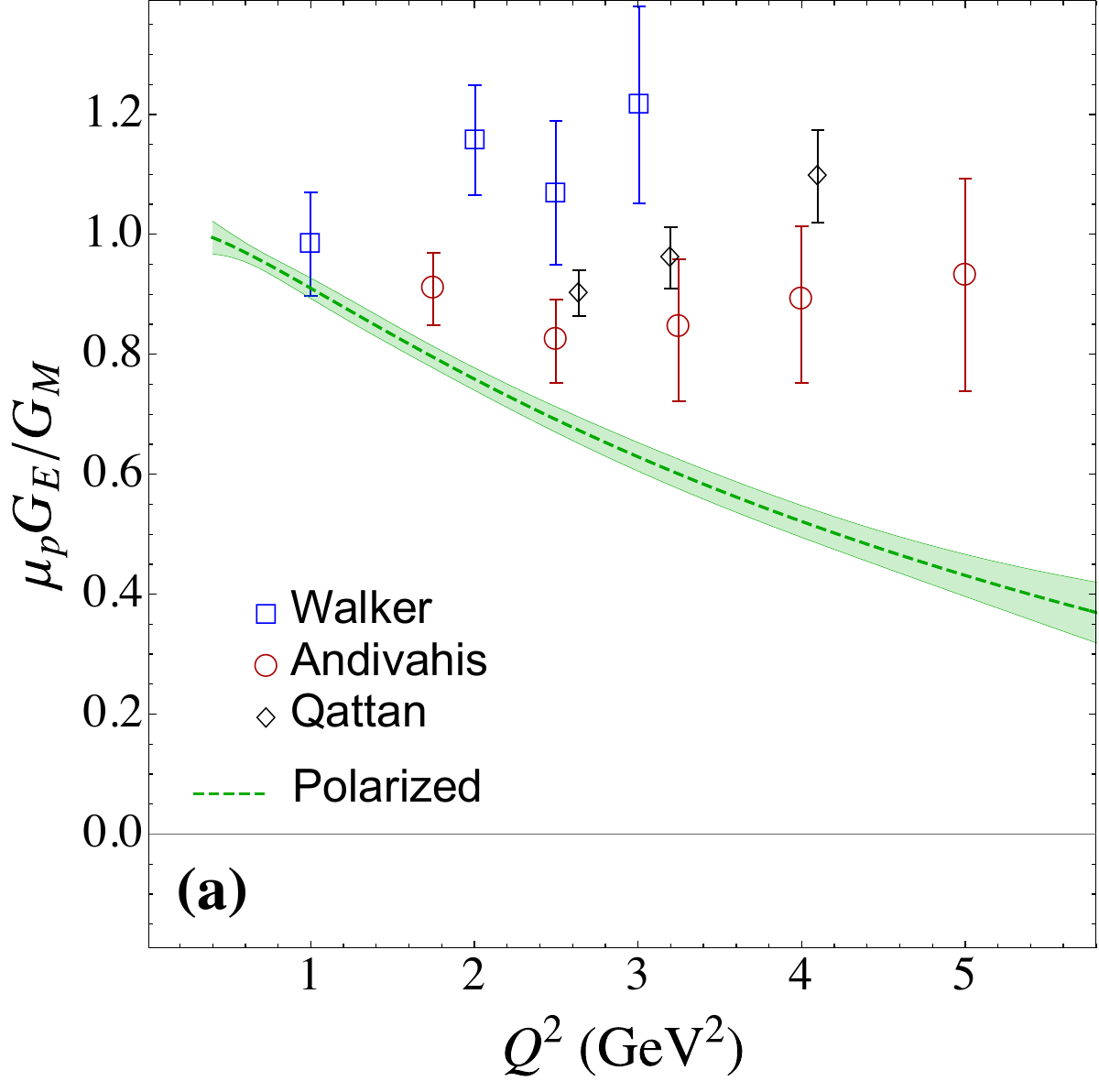}\\
\includegraphics[width=0.42\textwidth]{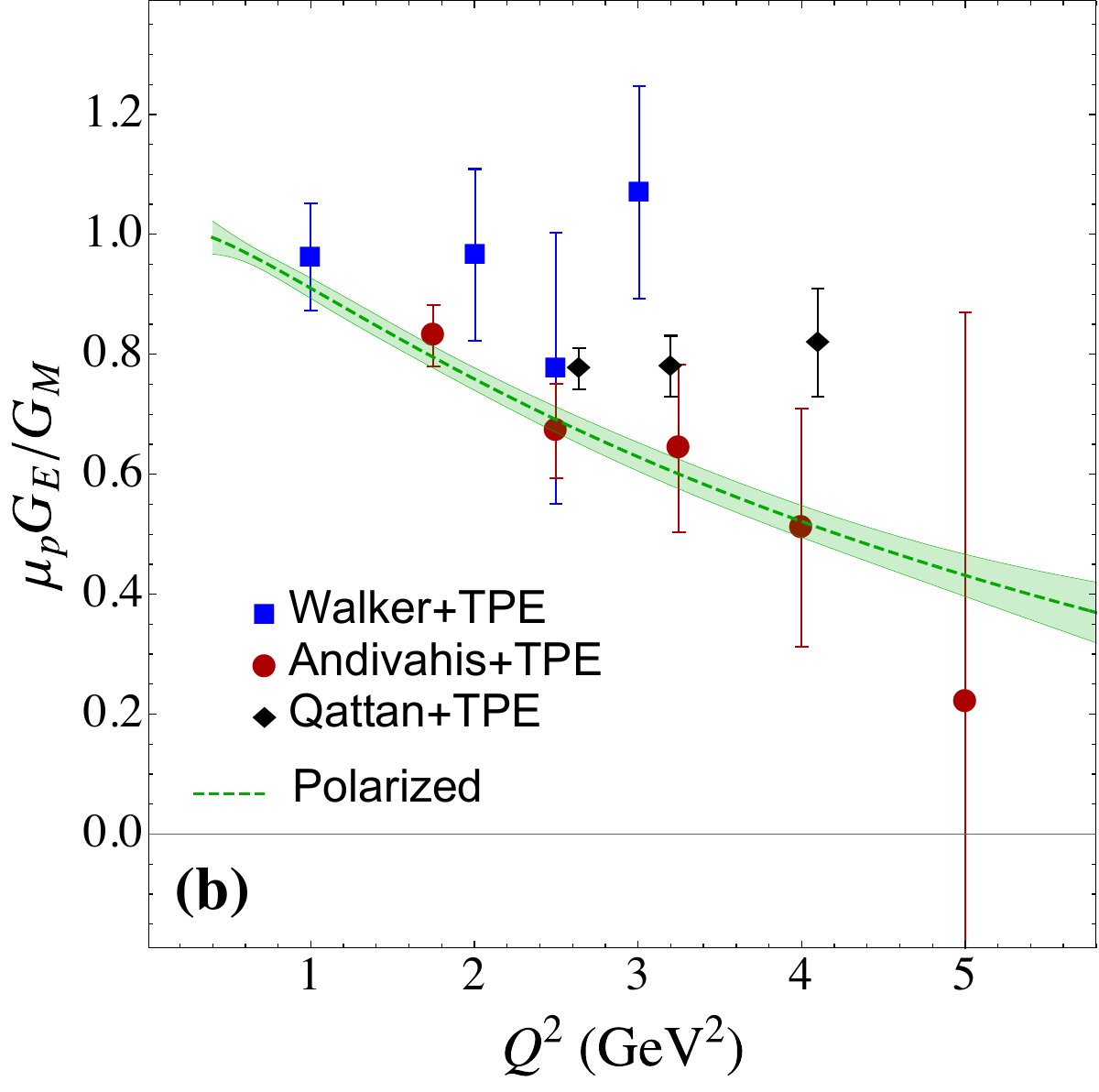}
\caption{(a) Ratio $\mu_p\, G_E/G_M$, versus $Q^2$, extracted using LT separation
data~\cite{walker1994, Andivahis1994, qattan2005}. A nonlinear fit to the
combined PT results~\cite{jones2000g, gayou2002, punjabi2005, puckett2010,
puckett2012} at the 99\% confidence limit is shown by the green band.
(b) The ratio $\mu_p\, G_E/G_M$ extracted from a reanalysis of the LT data
using improved standard RCs from Ref.~\cite{Gramolin2016}, together with the TPE
effects from our work~\cite{Ahmed2020}.}
\label{fig.GeGmRatio}
\end{figure}

In Fig.~\ref{fig.GeGmRatio} we show the $G_E/G_M$ ratio extracted from our
analysis for the SLAC~\cite{walker1994, Andivahis1994} and Jefferson Lab
Super-Rosenbluth~\cite{qattan2005} experiments up to $Q^2=5$~GeV$^2$. To avoid
clutter, the PT data from Refs.~\cite{jones2000g, gayou2002, punjabi2005,
puckett2010, puckett2012} are shown as a band, which is a nonlinear fit at the
99\% confidence limit. The original analysis, shown in
Fig.~\ref{fig.GeGmRatio}(a), is consistent with $\mu_p\, G_E/G_M \approx 1$, while
a progressively larger effect of TPE with increasing $Q^2$ for all LT data sets
is seen in Fig.~\ref{fig.GeGmRatio}(b), with a commensurate increase in the
uncertainty of $G_E$. 
In particular, the LT data of Andivahis~{\it et al.}~\cite{Andivahis1994}
are striking in their consistency with the PT band, with a near linear falloff of
$G_E/G_M$ with $Q^2$. These results provide compelling evidence that there is no
inconsistency between the LT and PT data once improvements in the RCs and TPE
effects are made.

Future precision measurements at higher $Q^2$ values and backward angles (small
$\varepsilon$), where  TPE effects are expected to be most significant, would
be helpful for better constraining the TPE calculations. This would provide a
more complete understanding of the relevance of TPE in the resolution of the
proton's $G_E/G_M$ form factor ratio puzzle, and better elucidate the role of
multi-photon effects in electron scattering in general.

\ctitle{Radiative corrections and two-photon exchange for MUSE and JLAB}
\byline{Andrei Afanasev, Alexandr Ilyichev}
Development of consistent approaches to QED corrections in lepton-nucleon scattering is crucial for precision analysis of nucleon's structure. While at JLab energies ultra-relativistic approximation is well justified for GeV electrons (except for very forward scattering angles), for MUSE muons with momenta in 100s MeV/c, the lepton mass effects have to be treated without assuming the small-mass limit. MUSE experiment aims to compare electron-proton and muon-proton cross sections with per-cent accuracy and probe two-photon effects via measurements of a charge asymmetry in the scattering of leptons vs anti-leptons. In addition to purely kinematic mass effects, interaction dynamics for muons contains new contributions, as we discuss below.

\begin{figure}[b!]
\includegraphics[width=\columnwidth]{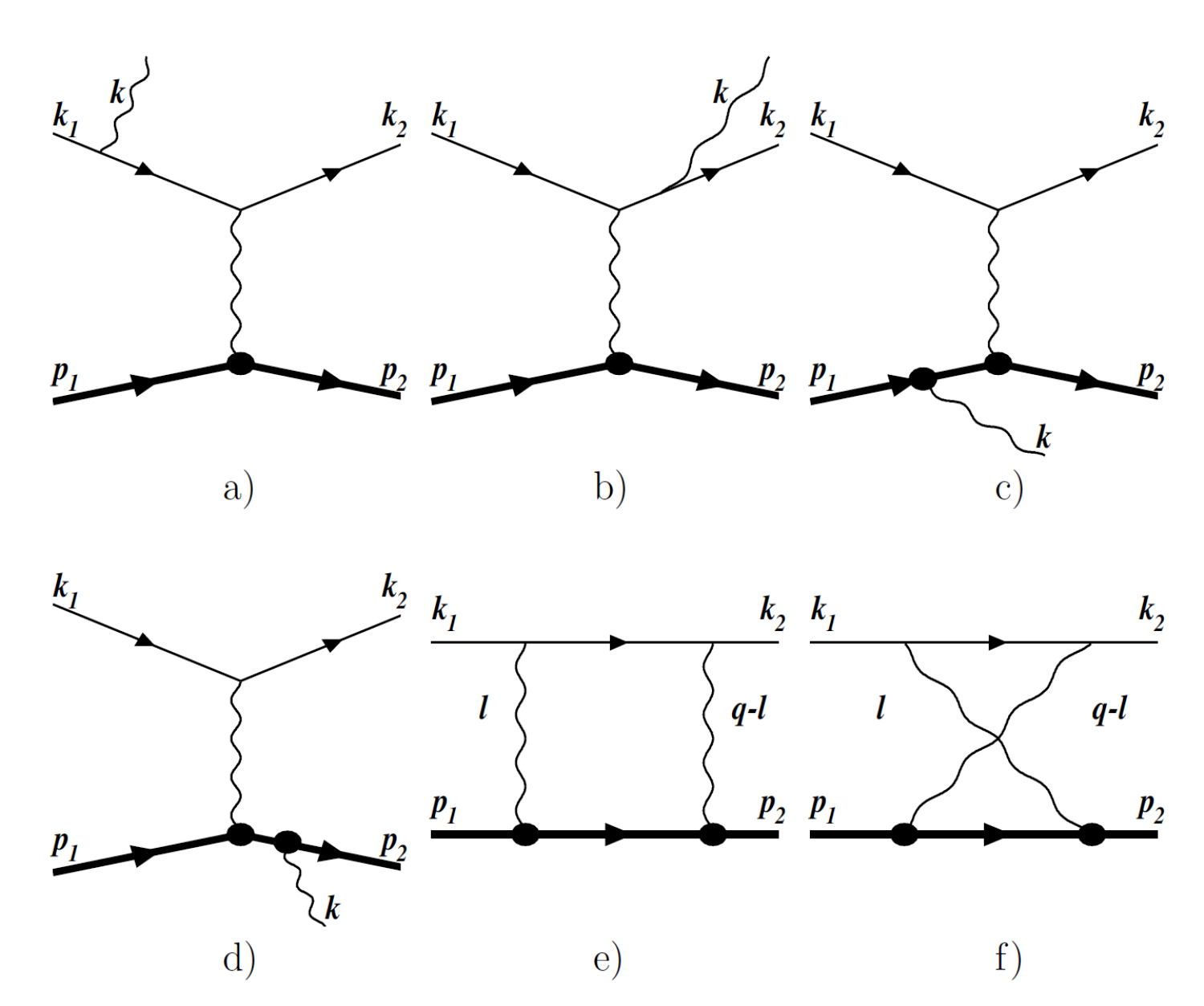}
\caption{Feynman diagrams for leading-order QED corrections responsible for charge asymmetries in lepton-proton scattering. The charge asymmetry is due to interference between one-photon exchange contribution and diagrams (c,d), as well as interference between photon emission from the lepton (a,b) and the proton.  }
\label{fig_aa_diagrams}
\end{figure}

Leading-order QED corrections to electron or muon scattering on a nucleon responsible for charge asymmetries are represented by the diagrams in Fig.\ \ref{fig_aa_diagrams}.

\begin{figure*}[t]
\centering
\includegraphics[width=\columnwidth]{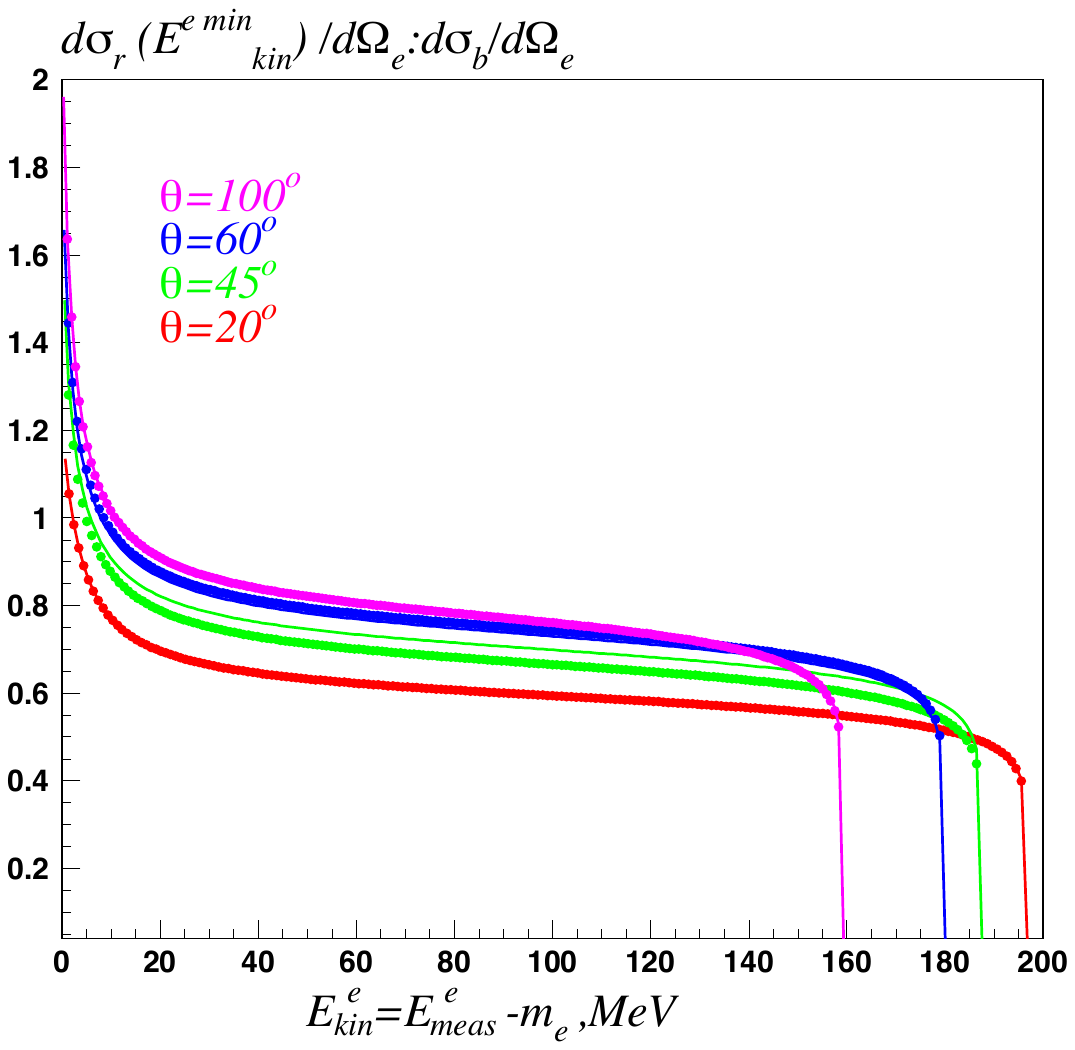}\hfill
\includegraphics[width=\columnwidth]{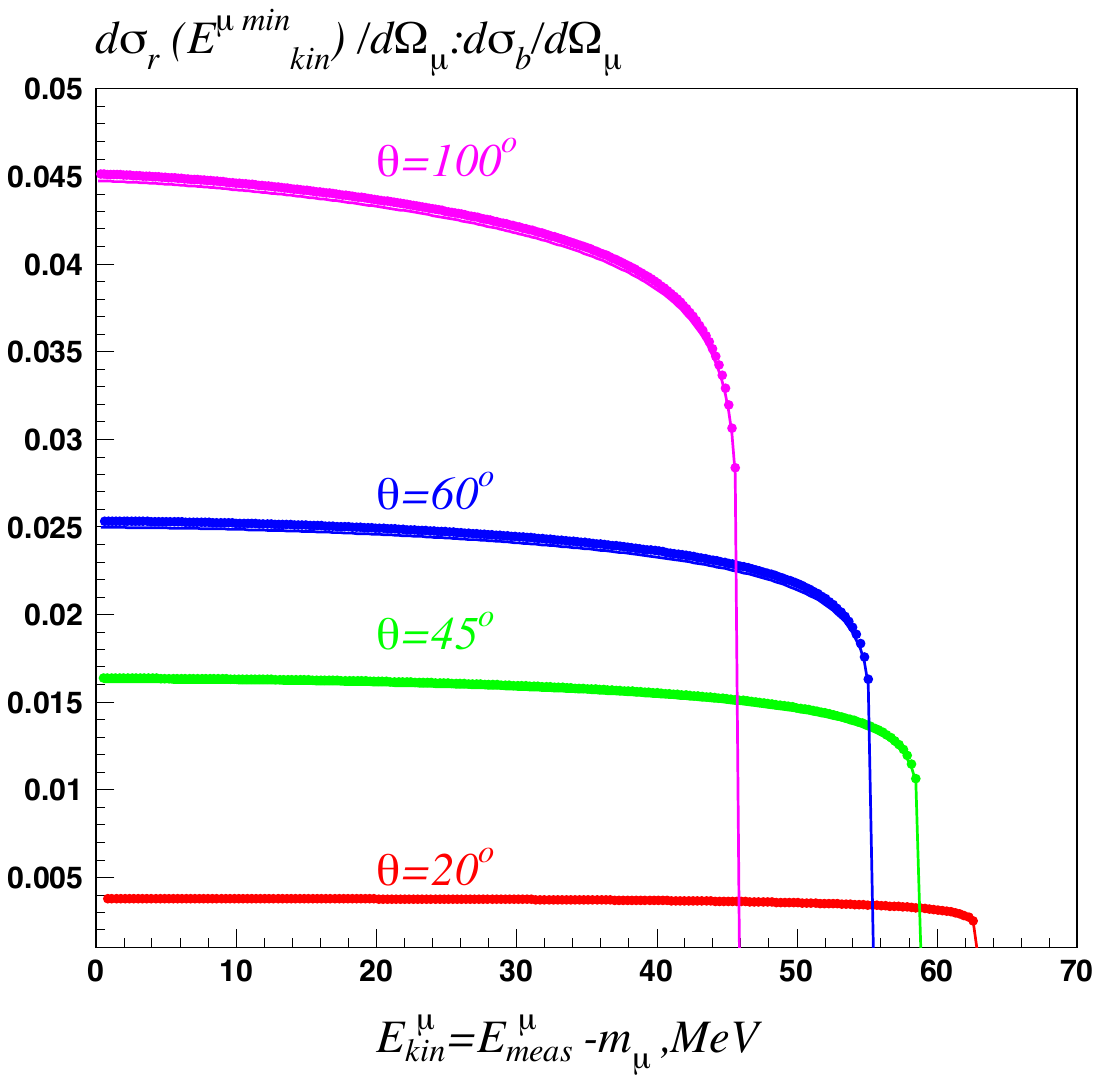}
\caption{QED correction to the cross section of electron-proton scattering (left plot) and muon-proton scattering (right plot) in MUSE kinematics as a function of final lepton energy plotted for different scattering angles. The correction for electrons larger and it is strongly peaked for nearly stopped final electrons, which makes it sensitive to energy cuts. The correction to muons is smaller, at one per cent level, and does not show drastic variation with final-muon energy.}
\label{fig_RC}
\end{figure*}

We have upgraded a Monte-Carlo generator ELRADGEN \cite{Afanasev:2003ic} to include lepton masses without using an ultra-relativistic approximation. QED corrections due to model-independent emission from leptons is shown in Fig.\ \ref{fig_RC}, calculated using ELRADGEN in MUSE kinematics. 
Considering only soft-photon corrections, the contributions of the diagrams Fig.\ \ref{fig_aa_diagrams} to scattering observables were calculated analytically in Ref.\cite{PhysRevD.96.016005} without using a small-mass approximation for the scattering leptons, and charge asymmetries were predicted for a wide range of kinematics, including MUSE. 

Among the dynamic effects different for muons and electrons beyond the leading order in QED, we should mention enhancement of helicity-flip amplitude for massive muons that makes a noticeable contribution to the two-photon correction  \cite{PhysRevD.94.116007} via $t$-channel $\sigma$-meson exchange; and a single-spin beam asymmetry for polarized muons (coming from weak decay of pions) that generates an azimuthal dependence of scattering cross section \cite{PhysRevD.100.096020}.
Two-photon corrections to muon-proton scattering were also considered in Ref.\cite{PhysRevD.90.013006}.

The next step was made recently \cite{Afanasev:2020ejr} with inclusion of the emission of {\it hard} photons to describe charge asymmetries in scattering of electrons and muons on a proton. While hard-photon emission may be constrained by kinematic cuts in experimental analysis, its inclusion is necessary for complete analysis, especially the events located further away from the elastic peak on a radiative tail. 

In summary, the main contributions to QED corrections for lepton-proton scattering were calculated and implemented in Monte-Carlo code ELRADGEN. The code can be used for MUSE kinematics, where ultra-relativistic approximation is no longer applicable for the muons, as well as at JLab and other facilities, where lepton mass effects play a role at very forward scattering angles (e.g., in PRAD experiment at JLab). Two-photon effects are included in a soft-photon-exchange approximation, while keeping options to include additional model-dependent two-photon contributions. The code can be downloaded from http://www.jlab.org/RC. 
\newpage

\ctitle{Radiative corrections for the MUSE experiment}
\byline{Lin Li, Steffen Strauch}

\section{Muon Scattering Experiment (MUSE)}
The MUon Scattering Experiment (MUSE) at the Paul Scherrer Institute (PSI)~\cite{Gilman:2017hdr} has been developed to measure elastic electron-proton and muon-proton scattering cross-sections of positively and negatively charged leptons and beam momenta between 115 MeV/$c$ and 210 MeV/$c$ over a wide angular range.  MUSE covers a four-momentum-transfer range from $Q^2 = 0.002$ to 0.08 GeV$^2$. Each of the four sets of data will allow the extraction of the proton charge radius. In combination, the data test possible differences between the electron and muon interactions and, additionally, two-photon exchange effects. 

\begin{figure}[b]
  \centering
  \includegraphics[width=0.9\columnwidth]{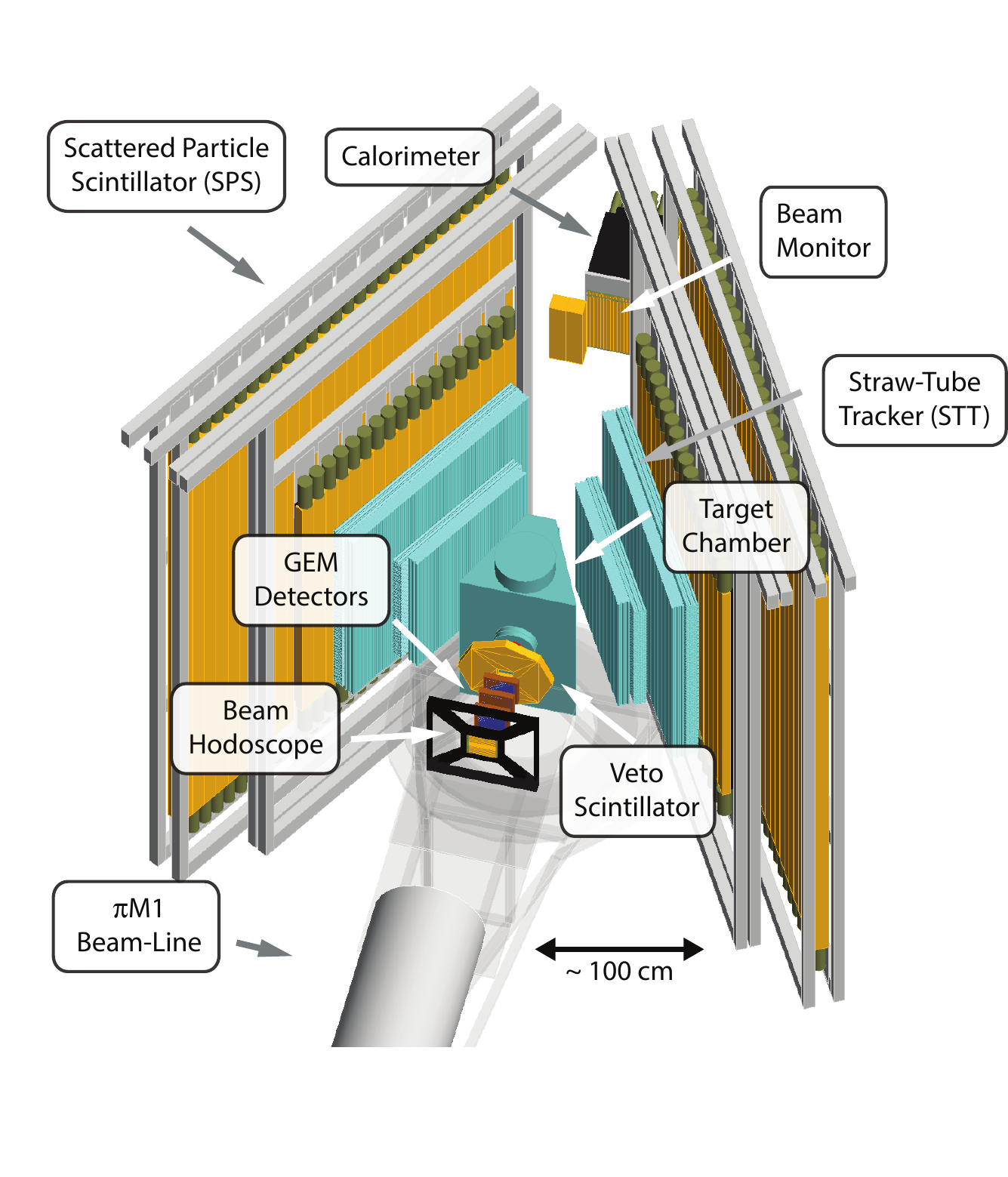}
\caption{Sketch of the MUSE experimental setup at the $\pi$M1 beamline at PSI as implemented in the MUSE Geant4 simulation.  The target chamber contains liquid-hydrogen, empty, and solid targets.}
\label{fig:muse_timing_2020920}
\end{figure}

\begin{figure*}[t]
 \centering
  \includegraphics[width=\columnwidth]{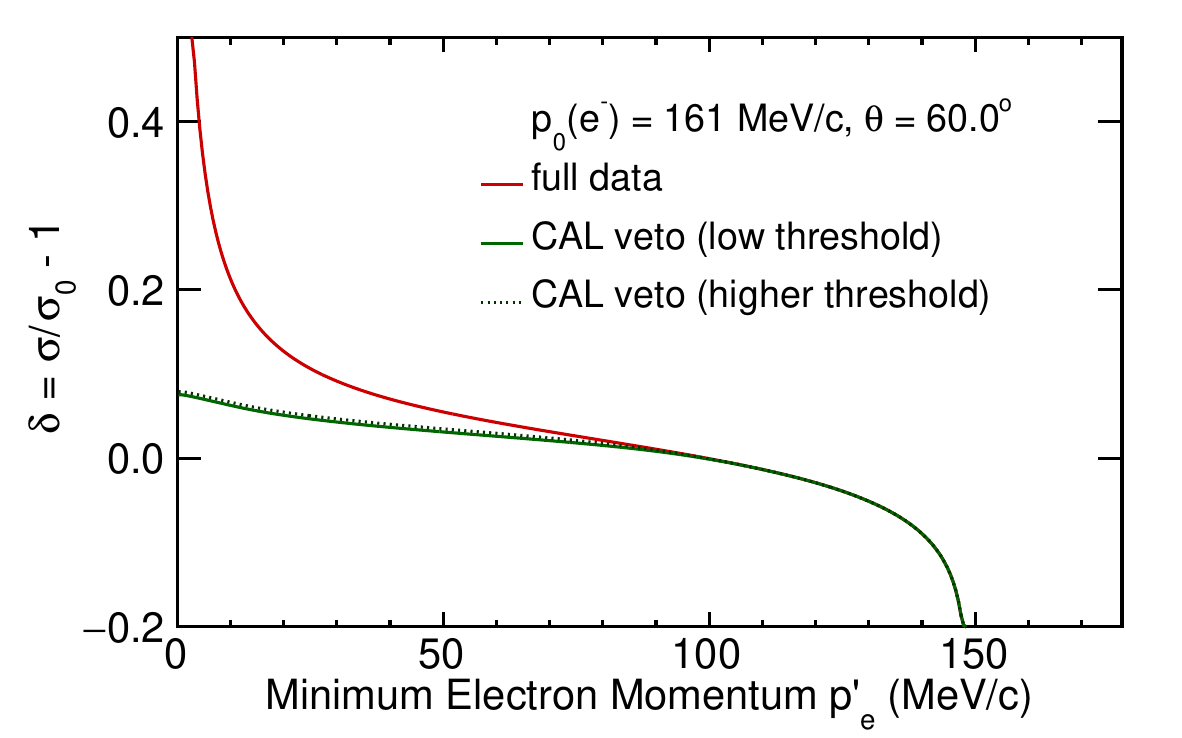}\hfill
  \includegraphics[width=\columnwidth]{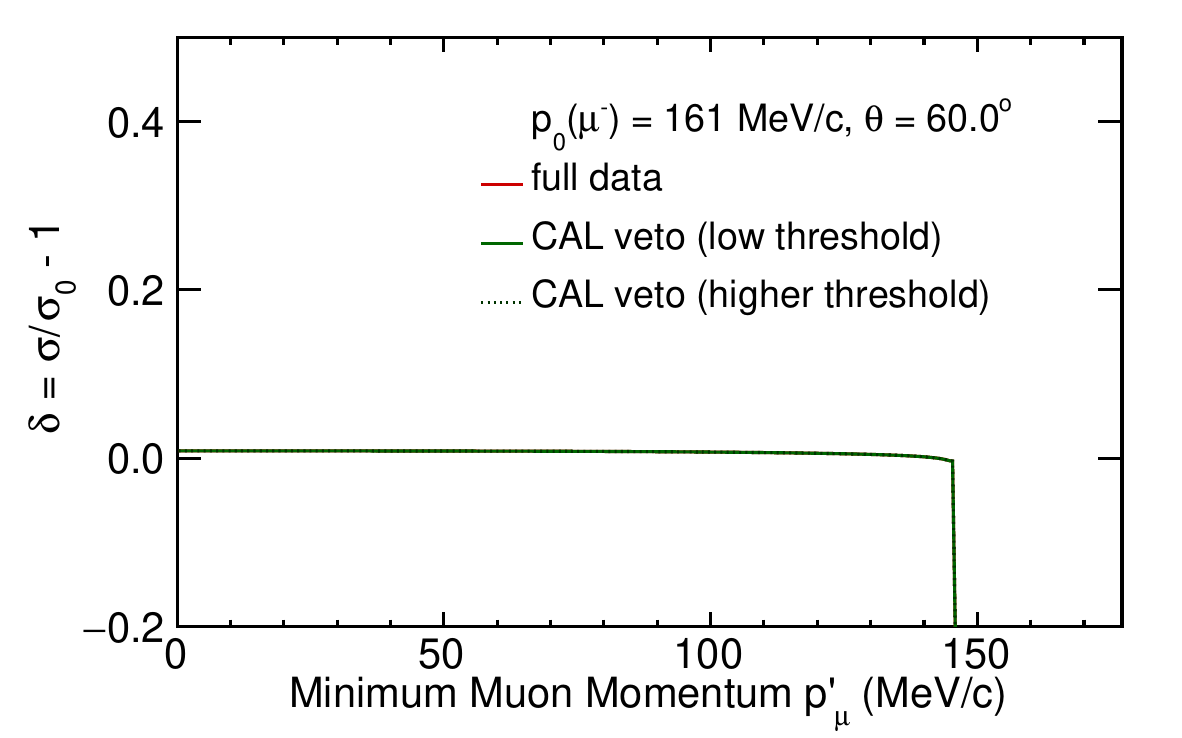}
  \caption{Preliminary radiative corrections for $ep$ (left) and $\mu p$ (right) scattering in typical MUSE kinematics. The red curve shows the full result without a photon veto. The green cuves are after suppression of initial-state radiation and include a veto on events with a hit in the calorimeter at different thresholds.}
\label{fig:delta} 
\end{figure*}

Figure~\ref{fig:muse_timing_2020920} shows a sketch of the experimental setup of MUSE at the secondary $\pi$M1 beamline at PSI. The particle beam contains a mix of electrons, muons, and pions. These particles go through the Beam Hodoscope (BH) detector for timing and particle identification. The BH also measures the beam fluxes.  Three Gas Electron Multiplier (GEM) chambers determine the incident particle track.  A veto detector suppresses triggers from off-axis particles. Particles that scatter from the Liquid Hydrogen target are detected by two symmetric spectrometers, each with two Straw-Tube Tracker (STT) chambers and two planes of fast Scattered Particle Scintillators (SPS). The two SPS planes (front-wall and rear-wall) provide the event trigger which requires an above-threshold ($E_{th}$ = 2 MeV) hit in each scintillator wall. The threshold for detecting electrons is around 10 MeV/$c$. The unscattered particles go through the Beam Monitor (BM). The time-of-flight (TOF) from BH to SPS determines the reaction type (muon scattering vs. muon decay in flight). TOF from BH to BM identifies backgrounds and determines $\mu$ and $\pi$ beam momenta. The BM can also be used to suppress M\o ller events. Downstream of BM is a calorimeter which is used to detect photons radiated from particles. 

\section{Radiative Corrections for MUSE}
For each bin of the scattering angle, MUSE will measure the integrated electron-proton and muon-proton yields that includes all events in the spectrometer acceptance and momenta above the detection threshold of $p^{'}_\text{min} \approx~10~\text{MeV/c}$. The magnitude of the momenta of the final-state leptons remain otherwise unmeasured. The Born cross-section, that contains information about the electromagnetic form factors, is obtained from the experimental cross-section, $(d\sigma/d\Omega)$, with a correction, $1 +\delta$, for higher-order processes, including the first-order bremsstrahlung process, the vacuum polarization correction, the vertex corrections, and the two-photon-exchange corrections, e.g., \cite{Maximon:2000hm},

\begin{equation}
\left(\frac{d\sigma}{d\Omega}\right)=\left(\frac{d\sigma}{d\Omega}\right)_{\text{Born}}(1+\delta)\;.
\end{equation}

Besides the internal processes where a photon is emitted during the lepton-proton scattering, the external processes accompanying the passage of incident and outgoing particles through the upstream detector and target materials have to be considered. To study the full effect, a Monte Carlo simulation with an event generator is being implemented for MUSE. An event generator for MUSE must fulfill several requirements. First, the event generator should include the emission of a hard radiated photon, which is beyond the soft-photon approximation. The hard photon needs to be propagated in the full Monte Carlo simulation. Second, the event generator needs to include the mass of the leptons, avoiding the $Q^2 \gg m^2$ approximation in the calculation. The Elastic Scattering of Electrons and Positrons on Protons (ESEPP) event generator \cite{Gramolin:2014pva} takes into account the first-order radiative corrections of elastic scattering of charged leptons ($e$$^{\pm}$ and $\mu$$^{\pm}$) on protons and fulfills these requirements. We have used ESEPP for our preliminary studies.  Other generators that are viable for MUSE include those of Refs.~\cite{Akushevich:2011zy, bernauer:thesis, Gramolin:2014pva, Talukdar:2018hia}.

The internal radiative correction $\delta$ depends on the kinematics of the reaction and detector properties.  Uncertainties in the knowledge of these parameters propagate into uncertainties in the corrections.  Using ESEPP, we studied the dependence of $\delta$ on the beam momentum, scattering angle, and the minimum lepton momentum $p'_\text{min}$, and found the largest sensitivity due to $p'_\text{min}$. Preliminary results for $\delta$ are shown in Fig.~\ref{fig:delta} as a function of $p'_\text{min}$ for electron (left panel) and muon (right panel) beams of a momentum of 161~MeV/$c$ and for a mid-range scattering angle of $60^\circ$.  The red curves show the results for all scattering events, regardless of the emission of photons.  It shows a strong dependence on $p'_\text{min}$ with a steep slope ($>1\%$ change per 1~MeV/$c$) close to the SPS detection threshold of 10~MeV/$c$.  If uncontrolled, the uncertainty in $p'_\text{min}$ would generate a considerable uncertainty in $\delta$.  The simulations reveal that the increase in $\delta$ with a decrease in $p'_\text{min}$ is linked to the increase of the $ep$ cross-section with reduced beam momentum after the emission of high-energy initial-state radiation.  The initial-state radiation is strongly forward peaked, and the MUSE calorimeter in the beamline downstream of the target is capable of detecting these Bremsstrahl photons.  The green curves in Fig.~\ref{fig:delta} show the results for $\delta$ after vetoing those events with a detected photon.  The difference between the solid and dashed green curves is in different photon-detection thresholds in the calorimeter. The radiative corrections are smaller after the suppression of initial-state radiation, and the dependence on  $p'_\text{min}$ is reduced. Radiative corrections for muons are much smaller than for electrons and nearly independent on $p'_\text{min}$, as shown in the right panel of Fig.~\ref{fig:delta}. Because there are not many photons emitting from muons to the forward direction, the calorimeter cut does not affect the data. 

Considering all the kinematic settings, the preliminary estimates of the uncertainties in the radiative corrections are 0.4\% to 0.8\% for electrons and smaller than 0.06$\%$ for muons.

\section{Summary and Outlook}
MUSE is a high-precision experiment to measure the proton charge radius, study possible 2$\gamma$ mechanisms, and have a direct $\mu / e$ comparison of the elastic cross-sections. Without a magnetic spectrometer, MUSE does not measure the final-state lepton momentum precisely. A dedicated downstream photon detector helps to suppress initial-state radiation effects. Preliminary ESEPP simulations show uncertainties in the radiative corrections for electron scattering to be lower than 1$\%$. Ongoing work includes implementing the event generator in a full MUSE simulation.  After data taking, the analysis allows for moderate changes in the SPS lepton-momentum detection threshold.  Those changes will affect radiative corrections. Consistent cross-sections extractions for various values of $p'_\text{min}$ test the validity of the corrections.

\ctitle{Modeling Radiative Processes for the OLYMPUS Experiment}
\byline{Axel Schmidt}

The OLYMPUS Experiment~\cite{Henderson:2016dea} was one of three recent experiments to quantify the effects of hard two-photon exchange (TPE) in elastic electron-proton scattering~\cite{Henderson:2016dea,Rachek:2014fam,Adikaram:2014ykv}. 
OLYMPUS measured the ratio of positron-proton to electron-proton elastic scattering cross sections, a ratio where deviations from unity are evidence of TPE. 
The goal of these experiments was to test whether the effects of hard TPE are sufficient to explain the discrepancy between polarized and unpolarized measurements of the proton form factor ratio $\mu G_E/G_M$.
Hard TPE is a radiative correction that has been typically assumed to be negligible, and one reason is that it cannot be calculated in a model independent way. 
By contrast, soft TPE---calculated in the limit in which one exchanged photon carries negligible momentum---is included in standard radiative correction formulae~\cite{Mo:1968cg,Maximon:2000hm}, though with slight differences in definition. 

\begin{figure}[htpb]
    \centering
    \includegraphics[width=\columnwidth]{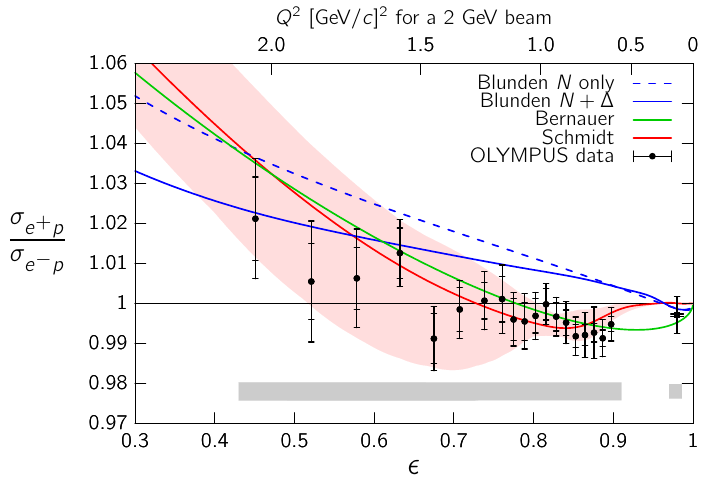}
    \caption{The OLYMPUS results~\cite{Henderson:2016dea} compared with the theoretical predictions of Blunden and Melnitchouk~\cite{Blunden:2017nby}, and the phenomenological predictions of Bernauer et al.~\cite{Bernauer:2013tpr} and Schmidt~\cite{Schmidt:2019vpr}. Though the data don't show large deviations from unity, the results are consistent with the size of the proton form factor discrepancy in the kinematics probed.}
    \label{fig:olympus_results}
\end{figure}

The results of the OLYMPUS Experiment are shown in Fig.~\ref{fig:olympus_results}, and compared to several predictions. The results show evidence of a small hard TPE effect. The data fall slightly below the dispersive hadronic calculations of Blunden and Melnitchouk~\cite{Blunden:2017nby}, and agree better with phenomenological predictions~\cite{Bernauer:2013tpr,Schmidt:2019vpr} made from the size of the proton form factor discrepancy. This suggests that even though the measured hard  TPE effect is small, it is fully consistent with the hypothesis that it is the cause of the form factor discrepancy. A more definitive measurement is needed at higher $Q^2$ / lower $\epsilon$. This would be particularly useful for testing partonic calculations of hard TPE (e.g., Refs.~\cite{Afanasev:2005mp,Kivel:2009eg}). The findings of these experiments and a summary of the current status of two-photon exchange can be found in Ref.~\cite{Afanasev:2017gsk}.

Accurately modeling radiative effects was important in OLYMPUS because isolating the effects of hard TPE meant also accounting for other radiative processes that introduce an asymmetry between positron-proton and electron-proton cross sections. 
In addition to soft TPE, the effects of interference between lepton-leg bremsstrahlung with proton-leg bremsstrahlung are also lepton-charge odd. These effects are treated in the standard radiative corrections formulae, but treating radiative corrections as a simple multiplicative factor was inappropriate for OLYMPUS for two reasons.
\begin{enumerate}
    \item OLYMPUS was a coincidence measurement, and elastic event selection was made on non-trivial exclusivity cuts.
    \item OLYMPUS did not have excellent momentum resolution, meaning that the elastic event sample had an unavoidable contribution from events with a hard radiated photon.
\end{enumerate}
Given this situation, we chose to model radiative effects using simulation and to write a custom event generator, described in Chap 5 of Ref.~\cite{Schmidt:2017qvu}. We then used the generator to integrate the radiative elastic cross section over the phase space defined by the elastic event selection criteria. 

One major design question we had to confront was whether or not to use exponentiation to attempt to treat radiation beyond the $\alpha^3$ order. Exponentiation can be used to treat radiation to all orders in the soft limit~\cite{Yennie:1961ad}. However, the limits of the OLYMPUS detector resolution required us to model radiation in the deep tail. 
We decided to try both exponentiated and non-exponentiated approaches and to compare results. 

\section{Non-Exponentiated Approach}

In the non-exponentiated approach, we set an arbitrary cut-off energy (with a value much smaller than the detector resolution, typically about 1~MeV) below which the event was considered to be ``near-elastic'' and above which the event was considered part of the ``hard tail.'' In the near-elastic regime, a standard multiplicative correction was used, i.e.,
\begin{equation*}
    \frac{d\sigma}{d\Omega} = \frac{d\sigma}{d\Omega}_\text{Born} \big(1+\delta(\Delta E_\text{cut-off})\big).
\end{equation*}
In the hard tail, the cross section was assumed to be the tree-level bremsstrahlung cross section, taking care to include all of the interferences terms of all of the diagrams. The non-exponentiated approach essentially used the same cross section model as employed in the ESEPP Generator developed for the Novosibirsk two-photon experiment~\cite{Gramolin:2014pva}.

\section{Exponentiated Approach}

The exponentiated approach was developed from an earlier generator produced by the Mainz A1 collaboration (and described in Ref.~\cite{bernauer:thesis}). While the phase space for electron-proton scattering with multiple undetected hard bremsstrahlung photons is six-dimensional (three degrees of freedom for the electron momentum, three degrees of freedom for the proton momentum), we first made the simplifying assumption that the kinematics were well-described by the five-dimensions phase space in which only one radiated photon is hard. Next, we assumed that radiated cross section could take an exponential form, i.e.,
\begin{equation}
    d^5\sigma = \frac{d \sigma}{d \Omega}_\text{Born} e^\delta ( \partial_{\vec{\gamma}} \delta ),
\end{equation}
where $( \partial_{\vec{\gamma}} \delta )$ is the differential change cross section with respect to the momentum of the hard photon. We assumed that this could be approximated by the ratio of the tree-level bremsstrahlung cross section to the elastic $ep$ cross section, and that the $e^\delta$ correction factor could be taken from a standard radiative correction. The final model cross section was given by:
\begin{equation}
    d^5\sigma = e^\delta \times d^5\sigma_\text{brems.}.
\end{equation}

\section{Results}

\begin{figure}[htpb]
    \centering
\includegraphics[width=\columnwidth]{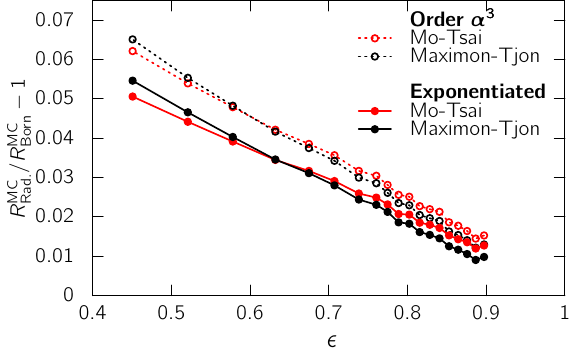}
    \caption{The charge-odd radiative corrections in OLYMPUS were several times bigger than the hard TPE effect. }
    \label{fig:rc}
\end{figure}

The approximate size of the charge-odd radiative correction in OLYMPUS  is shown in Fig.~\ref{fig:rc} for both the exponentiated and non-exponentiated (Order $\alpha^3$) approaches, for the different definitions of soft TPE employed by Mo and Tsai~\cite{Mo:1968cg} and Maximon and Tjon~\cite{Maximon:2000hm}. In both approaches, the correction grows as $\epsilon$ decreases, reaching 5--7\%, i.e., the measured hard TPE effect sat on top of a baseline ratio that was 5--7\% higher than unity due to soft TPE and bremsstrahlung interference. These radiative effects were several times bigger than the effect of hard TPE itself. The exponentiated approach predicted a smaller correction than the non-exponentiated approach; however, the difference was relatively small compared to the magnitude of the correction.

The OLYMPUS results were published relative to all four corrections, to allow the reader to judge based on the correction they deem most appropriate~\cite{Henderson:2016dea}. 

\section{Future Plans}

There are two obvious things that could be added to the OLYMPUS generator. First is the inclusion of models of hard TPE, in order to see how these effects would propagate through the entire simulation. Second is the inclusion of off-shell currents when calculating the matrix elements for bremsstrahlung emitted from the proton. So far, the OLYMPUS generator assumes on-shell currents even for these off-shell vertices. This would probably have a minute effect on the results, but is an easy improvement to make. 

OLYMPUS has recently submitted for publication the results of the lepton-charge average cross section (i.e. $\frac{1}{2}[e^+p + e^-p]$), a combination in which charge-odd effects like TPE cancel at lowest order~\cite{Bernauer:2020vue}. This observable can give direct information about the proton's form factors, while being much more robust to any bias from TPE. 

A paper describing the OLYMPUS generator in detail is in preparation and will hopefully be useful to future high-precision elastic scattering experiments, such as MUSE at PSI~\cite{Gilman:2017hdr}, the prospective TPEX project at DESY, or future experiments suggested by the Jefferson Lab positron working group~\cite{Accardi:2020swt}.

\ctitle{Radiative correction studies for the PRad and proposed PRad-II, DRad and SoLID experiments at Jefferson Laboratory}
\byline{Vladimir Khachatryan, Duane Byer, Haiyan Gao, Chao Peng, Weizhi Xiong}

\section{Radiative correction studies for the PRad experiment}
In order to reach a high precision in proton (deuteron) radius measurements from electron scattering, in addition to a tight control of 
various experimental systematic uncertainties, a careful calculation of radiative corrections (RC) is necessary.  A recent measurement 
of the charge radius of the proton was performed by the PRad experiment, which measured the unpolarized elastic $e+p$ 
scattering cross section and the proton electric form factor in an unprecedentedly low momentum transfer squared region: 
$Q^{2} = 2.1 \cdot 10^{-4} - 6 \cdot 10^{-2}~{\rm (GeV/c)}^{2}$, equivalent to the scattering angle in the laboratory frame, 
$0.8^{\circ} \leq \theta \leq 7^{\circ}$. During the PRad experiment the luminosity was monitored by simultaneously measuring the 
M{\o}ller scattering process, and the absolute $e-p$ elastic scattering cross section was normalized to that of the M{\o}ller to cancel out 
the luminosity. Therefore, the RC effects have been studied for both scattering processes. The extracted value of the proton radius turned 
out to be $r_{p} = 0.831 \pm 0.007_{\rm stat} \pm 0.012_{\rm syst}$~fm \cite{Xiong:2019umf,Xiong:2020}, with the RC being one of the 
largest systematic uncertainty sources.

Ref.~\cite{Akushevich:2015toa} shows a complete set of analytical expressions for calculating the one-loop RC diagrams to 
$e-p$ and M{\o}ller scatterings, obtained within a covariant formalism and beyond the ultrarelativistic approximation, where the infrared
divergence is extracted and canceled using the Bardin-Shumeiko approach \cite{Bardin:1976qa}. A related event generator has been 
built, which included RC contributions to the Born cross sections of these scattering processes \cite{PRadAnalyzer}. PRad has estimated 
the uncertainty on the measured proton radius based upon the first order RC results from \cite{Akushevich:2015toa}, and also using the 
method of \cite{Arbuzov:2015vba} for estimation of the contribution coming from higher order RC. The estimated systematic uncertainties 
for both $e-p$ and M{\o}ller scatterings are correlated and $Q^{2}$-dependent. The $Q^{2}$-dependence is much larger for the 
M{\o}ller RC, and it affects the cross section results through the use of the bin-by-bin method \cite{Xiong:2020}. If one transforms 
the cross section uncertainties into the uncertainties of $r_{p}$, then for $e-p$ we have $\sim 0.0020$~fm, and for the M{\o}ller we have 
$\sim 0.0065$\,fm, such that the total systematic uncertainty due to the higher order RC is equal to $\delta r_{p} = 0.0069$\,fm.

Given that the $Q^{2}$-dependent systematic uncertainty is much larger for the M{\o}ller scattering, PRad has performed an independent 
estimate for the second order RC effect on $r_{p}$. This estimate follows the method shown in \cite{Aleksejevs:2013gxa} (developed for the 
MOLLER experiment at JLab), where the authors have calculated two-loop electroweak corrections to the parity violating asymmetry to the 
M{\o}ller scattering in the MOLLER kinematic range. Based on their mathematical framework, we were able to estimate the contribution 
from the next-to-next leading order (NNLO) diagrams to the Born cross section in the PRad kinematic range (however, considering only
restricted number of diagrams). The estimated $Q^{2}$-dependent systematic uncertainties are smaller than those estimated in the first 
approach, for any reasonable photon energy cut for the PRad experiment. We obtained a few values of $\delta r_{p}$, with the largest one 
being 0.0047~fm, which is well within the first estimate. Thus, we still use the uncertainty of $\delta r_{p} = 0.0069$~fm due to RC effects 
as a conservative estimate on $r_{p}$ for PRad.

\section{Planned radiative correction studies for the proposed PRad-II experiment}
The proposed PRad-II experiment at JLab \cite{Gasparian:2020hog} -- an enhanced and improved version of the PRad experiment --
has a factor of 3.8 smaller estimated total uncertainty on $r_{p}$ than that of PRad. The PRad-II experiment will have projected total 
uncertainty of 0.43\%, by which one could address the possible systematic differences between $e-p$ and $\mu H$ experiments, as well as 
between the PRad and other recent $e-p$ scattering experiments. The improvements that PRad-II will achieve in reducing the key systematic 
uncertainties as compared to PRad,  includes in particular the installation of planes of high spatial-resolution tracking detector
based on the $\mu$RWELL, by which the detector efficiency determination will be much more precise. In the PRad experiment, one plane of
GEM  detectors was used for tracking. As it is shown in the PRad-II proposal, 
the $Q^{2}$-dependent systematic uncertainties from the M{\o}ller scattering can be suppressed by using the integrated M{\o}ller method for 
all angular bins \cite{Xiong:2020}, which will turn all systematic uncertainties from the  M{\o}ller into normalization uncertainties for the cross 
sections. This method is fully applicable for PRad-II due to presence of the two planes of tracking detectors, releasing us from the need 
of using the bin-by-bin method. Based on the studies for PRad-II, the RC uncertainty of $r_{p}$ coming from the integrated M{\o}ller method 
will be significantly reduced by a factor of $\sim 4$.

However, it would be very relevant to calculate the higher order RC contribution from the theory side as well. Along with other improvements 
in reducing the $r_{p}$ systematic uncertainties (see \cite{Gasparian:2020hog}), one more key factor is our planned new RC calculations to 
further improve the precision in determination of the proton radius. In this connection we should note that one of priority goals of PRad-II is to 
calculate exactly the next-to-next-leading (NNLO) RC diagrams in unpolarized elastic $e-p$ and M{\o}ller scatterings beyond ultrarelativistic limit 
for the PRad-II kinematics: namely, in the region of $Q^{2} \sim 10^{-5} - 6 \cdot 10^{-2}~{\rm (GeV/c)}^{2}$, corresponding to the scattering 
angle in the laboratory frame, $0.5^{\circ} \leq \theta \leq 7^{\circ}$. Currently our theory collaborators are developing a new method 
\cite{Srednyak:2020}, where the calculated results, namely amplitudes or cross sections, can be represented as power series that are convergent 
on the chain of integration of the corresponding integrals. With this method one may be able to calculate all necessary NNLO diagrams. The current 
PRad event generator \cite{PRadAnalyzer}, which is made based upon the NLO results from \cite{Akushevich:2015toa}, will be modified and 
developed accordingly, along with upcoming new results.

\section{Planned radiative correction studies for the proposed DRad experiment}
The PRad collaboration has also proposed the DRad experiment \cite{DRad} for the deuteron charge radius measurement at JLab, with recently 
estimated total uncertainty on extracted $r_{d}$ to be 0.22\%. In general, this experiment aims to accomplish a new high precision unpolarized 
elastic $e-d$ scattering cross section measurement in the low-$Q^2$ region of $2 \cdot 10^{-4} - 5 \cdot 10^{-2}\,{\rm (GeV/c)}^{2}$, using the 
PRad experimental setup with some modifications. All the M{\o}ller RC studies, which will be done for PRad-II, will be likewise applicable for DRad. 
However, for the studies of radiative effects in $e-d$ scattering, first we need to revise the papers \cite{Akushevich:1994dn,Akushevich:2001yp} 
and fulfill complete calculations of the NLO RC diagrams in unpolarized elastic $e-d$ scattering beyond ultrarelativistic limit for the DRad kinematics, 
by using the most recent charge, magnetic and quadrupole form factors of the deuteron. Accordingly, a new $e-d$ event generator will be created, 
which will replace the current existing one that is based on the soft photon approximation \cite{Gu:2017}.

\section{Ongoing radiative correction studies for SIDIS processes and making a related event generator for the proposed SoLID experiment}
In the upcoming years, highly accurate electron scattering experiments will be curried out in the 12 GeV Jlab program \cite{Dudek:2012vr}, which 
will provide unique opportunities for breakthrough studies of nucleon structure, based upon precision measurements of single and double spin 
asymmetries from semi-inclusive deep-inelastic-scattering (SIDIS) experiments to probe the partonic structure in momentum space, by utilizing 
detectors with high luminosity in combination with large acceptance. The Solenoidal Large Intensity Device (SoLID) will be such a high luminosity 
and large acceptance tool \cite{Chen:2014psa}, very well adapted for studies of nucleon structure. 

It is well known that one of the important sources of systematic uncertainties in SIDIS experiments is the QED RC. For example, the systematic 
uncertainties on asymmetry measurements for SIDIS in SoLID are estimated to be 2-3\% due to the RC effect. Currently, we are developing a new 
standalone event generator for polarized SIDIS including the finite mass of the lepton and QED RC following the calculations of 
Ref.~\cite{Akushevich:2019mbz}. This paper shows exact analytical expressions for the lowest order RC to SIDIS of polarized particles, obtained in 
the most compact, covariant form convenient for a numerical analysis. The calculations of RC have been performed in a model-independent way that 
involves constructing and using the SIDIS (and exclusive) hadronic tensor containing the eighteen SIDIS (and exclusive) structure functions. The SIDIS 
structure functions can be expressed in terms of Fourier-transformed transverse momentum-dependent parton distribution functions and fragmentation 
functions. The model-independent RC calculations in \cite{Akushevich:2019mbz} include the contributions of radiated SIDIS processes with loop 
diagrams, and the contribution of the exclusive radiative tail.

Configurability and extensibility are primary goals for the development of the aforementioned event generator, including providing either weighted or 
unweighted events, using approximations to the cross section for performance (such as the ultrarelativistic approximation), and using user-provided 
SIDIS structure functions. The generator software is being developed in parallel between a Mathematica prototype and the final C++ library. 
The C++ library will be developed with a focus on performance, particularly with regard to the evaluation of high-dimensional integrals from the SIDIS 
structure functions and partonic degrees of freedom. Pending further investigation, either the FOAM library or a Markov Chain Monte Carlo method
\cite{Guilhoto:MCMC} will be used for evaluating these integrals and providing unweighted events to the users. This new generator will be implemented 
into comprehensive SoLID SIDIS full simulations. \\

This work is supported in part by the U.S. Department of Energy under Contract No. DE-FG02-03ER41231.

\ctitle{Radiative corrections for MUSE}
\byline{F. Myhrer,  P. Talukdar, V. C. Shastry, U. Raha }
\title{Radiative Corrections for low-energy lepton-proton scattering in HB$\chi$PT }

\maketitle

\noindent 
{\bf Introduction} 
Determinations of the proton's r.m.s.\ radius 
have produced results which are not consistent with earlier results~\cite{
pohl2013muonic}. 
The proton radius puzzle refers to the contrasting results obtained between the proton's electric 
charge radii extracted from the Lamb shift of muonic hydrogen atoms 
and those extracted from electron-proton  scattering measurements (see, e.g., 
Refs.~\cite{Bernauer:2010wm,
Mihovilovic:2016rkr}.) 
In order to resolve this issue, a number of newly commissioned  experiments are underway, 
along with proposals of redoing the lepton proton  scattering measurements at low momentum transfers. 
The latter includes the MUon proton Scattering Experiment (MUSE)~\cite{Gilman:2013eiv,Gilman:2017hdr}. 
The MUSE collaboration proposes to measure the elastic differential cross sections for 
 $e^\pm p$ and $\mu^\pm p$ scattering at very low momentum transfers, and 
 aims for  an r.m.s.\ radius error of about 0.01 fm. 
%
However, the lepton beam momenta considered by MUSE are  of the order of the muon mass~\cite{Gilman:2013eiv,Gilman:2017hdr}. 
A particular concern is therefore the standard radiative correction procedure, see, e.g.,  
Refs,~\cite{
Maximon:2000hm,
Mo:1968cg}.       
The inelastic bremsstrahlung process is an integral part of the lepton-proton elastic scattering, 
and is a  sources of uncertainty for an accurate determination of the value of the momentum transfer $Q$.  
The influence of the in principle detectable 
"hard" photon bremsstrahlung spectrum is discussed in earlier publications, e.g.~\cite{Talukdar:2018hia} and will not be considered. 
We just note that in order to determine the proton charge radius, the data analysis necessarily need  
to correct for  this inelastic radiation process.  
We will concentrate on  the emitted very {\it soft} bremsstrahlung photons in order to evaluate the 
complete radiative corrections to the elastic lepton-proton scattering $\delta_{2\gamma}$.  
This radiative correction has been evaluated in a model independent way using {\it Heavy Baryon Chiral Perturbation Theory} ($\chi$PT) to 
next-to-leading order (NLO)\cite{Talukdar_2020}. 
We find the  observed differential cross-section to be 
\begin{eqnarray}
\left[\frac{{\rm d}\sigma^{\rm (NLO)}_{el} (Q^2)}{{\rm d}\Omega^\prime_l}\right]_\gamma=
\left[\frac{{\rm d}\sigma_{el} (Q^2)}{{\rm d}\Omega^\prime_l}\right]    
\left(1+\delta_{2\gamma}\right) 
\nonumber
\end{eqnarray} 
We note that the electron behaves relativistically whereas the muon mass is not ultra-relativistic at these low MUSE energies,
see Ref.~\cite{Talukdar_2020} for complete discussions.  

{\bf Radiative Corrections to Low Energy Lepton Scattering} 
At  low energy scattering  
hadrons are the relevant degrees of freedom, 
the dynamics is  governed by chiral symmetry requirements. 
$\chi$PT is a low-energy hadronic effective field theory (EFT), which incorporates the underlying symmetries 
and symmetry breaking patterns of QCD. 
In $\chi$PT the evaluation of  observables follow  well-defined chiral 
power counting rules, which  determines the dominant {\it leading order} (LO) contributions, as well as the 
NLO and higher order corrections  to observables in a  perturbative scheme. 
For example, in  $\chi$PT the proton's r.m.s. radius enters at a  chiral order where one encounters pion loops 
at the proton-photon vertex,  e.g., Ref.~\cite{Bernard:1995dp}. Furthermore, $\chi$PT               
naturally includes the photon-hadron coupling in a gauge invariant way. 

We treat the leptons relativistically, i.e. the 
lepton-current is 
    $ J^\mu_l(Q)=e \bar{u}_l(p^\prime )\gamma^\mu u_l(p)$, 
where the four-momentum transfer to the proton is $Q=p-p^\prime$, and the 
lepton mass $m_l$ is  included in all our expressions.    
The hadronic current is derived from the $\chi$PT Lagrangian. 
%
%
The proton mass, $m_p$, is large, of the order of the chiral scale $\Lambda_\chi \sim 4\pi f_\pi \sim 1$ GeV, 
where $f_\pi=93$ MeV is the pion decay constant.   
The expansion of the Lagrangian 
in powers of $m_p^{-1}$ is an  integral part of $\chi$PT.  
%
The {\it dynamical} $m_p^{-1}$ corrections arise from the NLO photon-proton interaction of the Lagrangian.    
In particular, the anomalous magnetic moments of the nucleons and the Pauli form factor enter in a natural way in 
the $\chi$PT formalism at NLO.   

We consider all photon loops in $\chi$PT including the {\it two-photon} exchange contributions. 
When necessary we invoke the {\it soft} photon approximation. Furthermore, the UV  and  IR  divergences 
are treated using the dimensional regularization scheme, 
see Refs.~\cite{Talukdar:2019dko,Talukdar_2020} for details. 
We find, for example, that the LO muon radiative correction goes through
zero at some low $Q$ value leaving the NLO corrections to be the dominant one around this particular $Q$ value. 
We remark that at low energy $\chi$PT we need a systematic improvement of the "radiative tail" following Ref.~\cite{Vanderhaeghen:2000ws}.

\ctitle{Radiative Corrections in SIDIS: Current Status and Perspectives}
\byline{Igor Akushevich and Alexander Ilyichev}
The model-independent radiative corrections (RC) to the semi-inclusive deep inelastic scattering (SIDIS) cross section contain four principal contributions: i) loop diagrams and soft photon radiation (indistinguishable because of the infrared divergence), ii) hard photon emission with semi-inclusive process, iii) hard photon emission with exclusive process, and iv) multiple soft photon emission. Each process is calculable exactly and the use of exact formulae in the RC procedure of experimental data with estimating systematic error due to RC is the gold standard in data analyses of modern experiments on lepton-nucleon scattering. By ``exact'' we understand the formulae allowing to exactly  reconstruct the leading and next-to-leading terms at least. Various approximations to exact formulae are possible but their accuracy can be estimated only based on comparison to the exact formulae. Our experience in analyses of data in deep inelastic scattering (DIS) tells us that the difference between estimates based on exact and approximate formulae can reach 100\% and strongly depends on a kinematical point. Possible (reasonable) approximations used in literature and data analysis practice include: i) soft photon approximation, ii) leading log approximation, iii) peaking approximation, iv) Compton-peak approximation, and v) using different Monte Carlo generators (e.g., RADGEN \cite{RADGEN}). The soft photon approximation is very convenient because RC is factorized at the Born cross section and completely cancels in spin asymmetries, but is wrong because of hard photon emission that cannot be neglected even in experiments with much poorer accuracies comparing to SIDIS experiments in JLab. The leading log (and peaking) approximation could estimate RC from hard photon with semi-inclusive process but is not applicable for the radiative tail from exclusive peak (or simply exclusive radiative tail). The Compton-peak approximation was good for the elastic radiative tail in DIS, but our estimates show that it will not work even for the exclusive radiative tail. 

The methodology used in experimental laboratories for accounting of radiative effects in SIDIS was basically a generalized version used for inclusive DIS studies, typically using RADGEN generator~\cite{RADGEN} combined with different full event generators, like PYTHIA\cite{Sjostrand:2006za}, LEPTO\cite{Ingelman:1996mq}, PEPSI \cite{Mankiewicz:1991dp}. That approach, while providing some estimates for RC, is inconsistent, as RADGEN itself contains only some version of DIS structure functions (SFs), and LUND base generators, completely ignore all kind of spin-orbit correlations. Unfortunately this approach is not applicable for analysis of azimuthal asymmetries in SIDIS, because key parts of the cross section responsible for $\cos(\phi_h)$ and $\cos(2\phi_h)$ do not appear in RADGEN. Therefore, the correct cross section of radiated photon cannot  be reconstructed using an approach involving pure-DIS Monte Carlo generators of RC. 
Precision studies of azimuthal moments in SIDIS will require a completely new methodology for accounting of radiative effects in SIDIS, taking as input some  set of realistic SFs describing all relevant moments for specific observables under study. The extraction of an azimuthal moment of interest, thus, will  require a full extraction of all azimuthal moments contributing to the SIDIS cross section for a given configuration of beam and target polarizations.

Therefore, the only way to provide a reasonable calculation of RC is based on the exact formulae for RC in SIDIS. Such formulae have to provide contributions from all 18 (5 spin-independent and 13 spin-dependent) SFs of SIDIS \cite{Kotzinian:1994dv} as well as all respective terms in the exclusive radiative tail.
Currently, the RC to the SIDIS cross sections and polarized asymmetries can be  calculated using POLRAD 2.0 \cite {Akushevich:1997di} and HAPRAD 2.0 \cite{Akushevich:2007jc}. In POLRAD 2.0, the RC cross section is calculated in terms of the parton distributions $f(x)$ and fragmentation functions $D(z)$, i.e., the simple quark-parton model is assumed. RC for unpolarized cross section and polarized part of the cross section are calculated, but no the exclusive radiative tail is separately calculated. Integration over $p_t$ and $\phi_h$ is assumed, i.e., the RC to the three-fold cross section is calculated: $d\sigma / dxdydz$. Several models for the parton distributions and fragmentation functions are implemented. In HAPRAD 2.0, RC
to the five-dimensional cross section
$ d^5\sigma/dxdydzdp_t^2d\phi_h$ are calculated. The contribution of  the exclusive radiative tail was implemented as a separate term that uses SFs for exclusive processes. Thus, although these codes are not applicable to polarized SIDIS or contain limitations that make them inapplicable in data analyses of modern SIDIS experiments, their use allows us to develop our  expectations on the size of the RC: i) $x$ and $Q^2$-dependences are similar to what we have in DIS; ii) RC go down with increasing $z$, e.g., RC factor can change from 1.05 to 0.85 between $z$=0.2 and 0.8 for the same $x$ and $Q^2$: the $z$-dependence of RC is generated by decreasing the phase space of radiated photon with increasing $z$; iii) $p_t$-dependence is strong: RC can increase by a factor of 2 or more for very high $p_t$: both semi-inclusive and exclusive processes have large RC for large $p_t$; iv) RC to $\phi_h$-dependence can be large (RC generate new $\phi_h$-dependence and therefore new observables like $ <\cos(3\phi_h)>$ that are exactly zero at the Born level); and v) RC from the exclusive radiative tail has its own dependence on kinematical variables and can give a high contribution especially as small $M_x^2$ (e.g., RC factor is 0.95 and 1.4 without and with the exclusive radiative tail for $M_x^2$=1.5 GeV$^2$ or 1.05 and 1.3 for  $M_x^2$=3.0 GeV$^2$) and for high $p_t$. Radiative corrections in the polarized case are largely unknown. From our experience in DIS, we expect similar patterns of RC and their larger size for the polarized part of the cross section. The effect is strongly dependent on the model for SFs. The strong model dependence can be partly addressed by applying the RC iteration procedure of experimental data. An illustration of possible effect of RC in unpolarized case shows that terms from SFs responsible for $\phi_h$-dependence can significantly contribute to the RC cross section with generation of higher harmonics that are significantly large in magnitude. For example, for kinematical point $E_{beam}$=10 GeV, $Q^2$=2.5~GeV$^2$, $x=0.25$, $z=0.2$, and $p_t$=0.7 GeV, the estimate of $<\cos(3\phi_h)>$ is about 5\%.

Recently, we completed the calculation of RC to SIDIS with longitudinally polarized lepton and arbitrary polarized nucleon \cite{Akushevich:2019mbz}. The exact calculation of RC for the SIDIS cross section contained contributions from all 18 SIDIS and 18 exclusive SFs and required derivation of the hadronic tensor for both the SIDIS and exclusive cross sections in the covariant form. Specifically, the 
stages of the theoretical calculation and practical implementation included the steps: i) elaborating the covariant hadronic tensor of SIDIS based on  developments of the theoretical groups of Aram Kotzinian \cite{Kotzinian:1994dv}, Peter Mulders \cite{Mulders} and Alessandro Bacchetta \cite{Bacchetta:2006tn}; ii) calculating the SIDIS cross sections using the elaborated hadronic tensor and compare analytic expressions for the Born cross section between the results obtained by researchers of these groups and understand possible discrepancies; iii) calculating the coefficients from convolution of the leptonic tensor involving RC with tensor structures from the hadronic tensor; and iv) evaluating the loop effects and extraction and cancellation of the infrared divergence using the covariant approach of Bardin and Shumeiko
\cite{BardinShumeiko}. The next steps could include implementation of the formulae to a code for numeric analyses, estimation of accuracies of most popular approximations, investigation of the model dependence of the RC, and identification of the kinematical regions where uncertainty in SFs could result in significant effect on RC. Note, any Monte Carlo generator that is required to reasonably simulate RC in SIDIS experiments must be based on the exact formulae. Therefore, we also intend to develop a Monte Carlo generator that will simulate the channel of scattering (non-radiative, radiative SIDIS, and radiative exclusive scattering) and the kinematical variables of the radiated photon. 
The strong model dependence can be partly addressed within
a realistic RC procedure of SIDIS experimental data that should involve an iteration procedure in which the fits of SFs of interest are re-estimated at each step of this iteration procedure. This procedure could be defined with and without involving Monte Carlo generator. Independently of whether MC is involved the procedure has to include the following steps: i)  
fitting the SIDIS SFs to have the model in the region covered by a SIDIS experiment; ii) using experimental data or theoretical models to construct the models in the regions of softer processes, resonance region, and exclusive scattering, iii) checking that the constructed models provide correct asymptotic behavior at the kinematical bounds (Regge limit, QCD limit); iv) jointing all the models to have continuous function of all four variables in all kinematical regions necessary for RC calculation; v) implementing this scheme in a computer code and defining an iteration procedure; vi) implementing the procedure of separation SFs if several SFs are measured in an experiment; vii) constructing the models for other SFs are if necessary (e.g., unpolarized SFs when spin asymmetries are measured); viii) paying specific attention to exclusive SFs, because the exclusive radiative tail is important (or even dominate) in certain kinematical regions; and ix) paying specific attention to $p_t$ dependence because RC is too sensitive for $p_t$ model choice.

In summary, newly achieved accuracies in Jlab and new physics studied at Jlab require paying renewed attention to RC calculations and their implementation in data analysis software. Although the theoretical calculation of RC to SIDIS is performed, current codes for computation of RC to SIDIS are not applicable to polarized SIDIS or contain limitations that make them inapplicable in data analyses of modern experiments. Lack of reasonable models for all 18 SIDIS SFs is the main obstacle in creating a workable code based on the exact formulae.  We expect sensitivity of the results for RC to specific assumptions used for constructing SIDIS SFs. Therefore, broad discussion and efforts of theoreticians and experimentalists are required to complete parameterizations of all SIDIS SFs as well as SFs in resonance region and exclusive SFs. The iteration procedure with fitting of measured SFs and joining with models beyond SIDIS measurements at each iteration step looks the better solution for specification of RC procedure of SIDIS experimental data.

\ctitle{New Physics at Low Energies and the Conceivable X17 Boson:
Perspective on Atomic Transitions and Scattering Experiments}
\byline{Ulrich D.~Jentschura}

{\em Introduction.---}At the moment, it seems that the proton 
radius puzzle could be solved by experimental considerations alone,
after new high-precision spectroscopic measurements have
revealed that the (certain hyperfine components of the) 
hydrogen $2S$--$4P$ frequency interval~\cite{Beyer:2017gug}
and the $2S$--$2P$ Lamb shift~\cite{Bezginov:2019mdi} are in agreement
with the new, lower value of the 
proton radius~\cite{Pohl:2010zza,Antognini:1900ns,Xiong:2019umf}.

On the other hand, 
the scattering experiment~\cite{Bernauer:2010wm},
which led to a roughly 5\% larger proton radius as compared to
Refs.~\cite{Xiong:2019umf}, constitutes an
exemplary accomplishment in terms of diligence in both
planning and execution. 
Furthermore, several high-precision 
spectroscopy measurements of the 
Paris group (e.g., Refs.~\cite{deBeauvoir:1997zz,Schwob:1999zz}
and also the new measurement~\cite{Fleurbaey:2018fih})
are in agreement with the ``larger'' proton 
radius.

In view of this interesting situation, we believe that,
even if the proton radius puzzle should be solved based
on a reconsideration of experimental aspects alone, 
one should keep in mind the possibility of the 
emergence of ``New Physics'' at the low-energy scale,
possibly visible in high-precision spectroscopy 
and scattering experiments.

One curious observation, recently made at the ATOMKI
Institute of Nuclear Physics in Debrecen, Hungary, concerns
the possible existence of a hitherto unobserved 17\,MeV 
gauge boson, termed the X17 boson~\cite{Krasznahorkay:2015iga,Krasznahorkay:2017gwn,Krasznahorkay:2019lyl},
which could modify scattering cross sections and high-precision spectroscopic 
measurements of simple atomic systems.

{\em The case for muonic bound systems.---}In two recent 
papers~\cite{Jentschura:2018zjv,Jentschura:2020zlr},
we have analyzed the possible effect of the X17 gauge boson 
on transition frequencies in simple atomic systems.
One of the general conclusions has been that in bound systems where the 
orbiting particle is an electron, the virtual X17 exchange
cannot easily be distinguished from nuclear effects, 
because the reduced Compton wavelength of the X17 is 
commensurate with the nuclear radius. We recall that
both the nuclear size (of the order of one fermi) as well as the 
reduced Compton wavelength of the X17 (of the order of about ten 
fermi, see Ref.~\cite{Jentschura:2018zjv}) are much smaller than the Bohr radius.

The situation is fundamentally different in muonic bound systems.
Namely, in muonic systems, the effective Bohr radius is smaller than in
electronic bound systems by roughly a factor of 100, 
and thus, the overlap of the 
X17-induced Yukawa-type potentials with the atomic 
wave function is much larger in bound systems where the 
orbiting particle is heavier than the electron.

It turns out that the effects induced by the X17 are 
still very elusive when it comes to measurements
of the Lamb shift~\cite{Jentschura:2018zjv}, 
even in muonic bound systems.
This is because of the small coupling parameters of 
the X17, which are much smaller than the QED fine-structure
constant, the latter being the coupling parameter of 
quantum electrodynamics (QED)
(see Refs.~\cite{Ellwanger:2016wfe,Feng:2016jff,Feng:2016ysn}). 

However, in the hyperfine splitting, the situation is different:
The magnetic field of the nucleus generates
the Fermi splitting~\cite{Itzykson:1980rh}, which, for $S$ states, is given
as the expectation value of a Hamiltonian
proportional to a Dirac-$\delta$ in coordinate 
space. From this observation alone, it is clear that 
hyperfine effects correspond to short-range
interactions, and that the effect of the X17 will be 
much more visible in hyperfine transitions as compared
to other transitions which involve
a change in the principle quantum number, or the 
angular-momentum quantum numbers ($\ell$ and $j$)
of the orbiting particle. 

According to Eqs.~(61a) and (61b) of Ref.~\cite{Jentschura:2020zlr},
the effect of the X17 enters the hyperfine splitting
of muonic deuterium already at the relative 
level of $10^{-6}$, expressed in terms of the
Fermi splitting. Hyperfine transitions 
among $P$ states suffer much less from uncertainties 
due to nuclear structure, due to the small
overlap of the $P$ state wave function with the 
nucleus. The effect of the X17 on $P$ states shifts the 
transition frequency by a relative fraction of 
a few parts in $10^{7}$. 
Under a moderate increase in the 
theoretical accuracy connected with nuclear-structure
effects, the X17 might thus be visible in muonic deuterium~\cite{Jentschura:2020zlr}.

Let us also consider
the ``purest and most interesting QED system imaginable'',
namely, true muonium (the bound system consisting of 
a muon and its antiparticle).
The designation is recalled from memory, 
from conversations with the 
late Professor Gerhard Soff~\cite{So1996priv},
who advocated and advised theoretical work on this 
system~\cite{Jentschura:1997tv,Karshenboim:1998we}.
In true muonium, the effects of the 
X17 on the hyperfine splitting are even more 
pronounced, entering the S state hyperfine splitting
at a few parts in $10^6$ [see Eq.~(69) of Ref.~\cite{Jentschura:2020zlr}]. 
A very moderate increase in the theoretical precision
for hadronic vacuum polarization effects~\cite{Lamm:2016vtf}
could thus make the X17 visible in 
true muonium.

{\em Implications for Scattering Experiments.---}We 
have singled out, above, the case of muonic deuterium
as opposed to hydrogen, because one of the two 
favored theoretical models for the 
X17~\cite{Feng:2016jff,Feng:2016ysn} predicts
a ``protophobic'' X17 which would, to a good approximation,
only couple to neutrons, not protons. 

In scattering experiments, one would thus expect the X17 to 
produce a slight ``halo'' in muon-deuteron scattering
experiments, which can be distinguished from the 
nuclear-size effect in the range of momentum transfers about 17 MeV
and larger.

For the other favored theoretical model of the X17 (Ref.~\cite{Ellwanger:2016wfe}),
which involves a conjectured pseudoscalar virtual particle,
the interaction is spin-dependent on the level 
of in the leading order
[see Eqs. (19) and (23c) of Ref.~\cite{Jentschura:2020zlr}],
and is thus visible, to good approximation,
only in the spin-resolved, differential cross 
section as opposed to the total scattering 
cross section.

These general observations on scattering experiments
would need to be substantiated 
by more concrete theoretical calculations and
feasibility studies. The effect of the X17 boson on scattering
experiments could provide grounds
for additional highly interesting theoretical investigations
and experiments.

{\em Proton size puzzle and new effects.---}On the occasion,
in addition to conceivable effects mediated by the X17 boson, 
we may discuss other conceivable new effects to 
be discerned in low-energy high-precision experiments.
Indeed, it may be premature 
at the current stage to completely 
discard other questions that surround the very recent, conflicting results
on the proton radius~\cite{Fleurbaey:2018fih,Xiong:2019umf}.

In particular, one might investigate the following question.
The proton wave function is an eigenstate of 
a Hamiltonian, which contains the sum of the quantized
electromagnetic interaction (quantum electrodynamics, QED)
and the quantized strong interaction (quantum chromodynamics, QCD).
The primary particle content of the proton is given by the 
three valence quarks. However, it is well known that 
the contribution of additional ``sea'' quarks also 
forms part of the proton wave function, in view of 
the nonperturbative character of the strong interaction.
Now, it can be checked by elementary calculations
that the electric field strength inside the proton
reaches values close to Schwinger's critical field
at which the vacuum ``sparks'' and electron-positron 
pairs are created. It is thus conceivable and in fact,
also inevitable~\cite{Buonocore:2020nai} that the proton wave function also contains
light ``sea'' fermions (primarily electron-positron pairs,
with muon pairs being kinematically suppressed in view of the 
larger mass).
The sea leptons are nonperturbative in the sense that they
belong to the initial state of the proton in the 
scattering experiment and thus are not comprised in the 
proton polarizability contribution to scattering, 
the latter having contributions from additional loops
with electron-positron pairs, which, however, have to 
be excited perturbatively during the scattering
(see Figs.~2 and 3 of Ref.~\cite{Jentschura:2014ila}).
If the light sea-fermion effect exists, then the proton radius 
would have to come out larger in electron scattering
than in muon scattering, because the virtual annihilation 
can only happen within one and the same particle 
generation. This conjecture would be consistent with 
the findings of a 1969 experiment~\cite{Camilleri:1969ag}.
The problem has been discussed further in 
Refs.~\cite{Miller:2015yga,Jentschura:2014yla}. With highly conflicting 
results on the proton radius~\cite{Fleurbaey:2018fih,Xiong:2019umf}
and with reference to 
the seasoned experiment~\cite{Camilleri:1969ag},
we believe that the proton radius puzzle will need to 
be explored further.

{\em Conclusions.---}For many decades, low-energy additions to the 
Standard Model have been explored without any imminent breakthroughs
in sight. High-precision 
spectroscopy and scattering experiments have looked for
such additions in tiny deviations of theory and experiment,
but so far, have found none. 
One should perhaps not forget that 
conjectures have been formulated for additions to the structure of the proton
(light sea fermions, Refs.~\cite{Jentschura:2014ila,Jentschura:2014yla})
which suggest that the proton radius puzzle could open the window
to new and interesting effects. 
At the moment, it could appear that the proton radius puzzle
is explained based on experimental advances alone~\cite{Beyer:2017gug,Bezginov:2019mdi,Xiong:2019umf},
but this finding would probably need to be confirmed
by additional experiments before a final line can be drawn.
In addition to increasing the accuracy of theoretical predictions
even further, a significant motivation for the 
experiments have been the hunt for tiny fifth-force interactions,
which could be visible in the 
deviations of theory and experiment.
The putative observation of the X17 
boson~\cite{Krasznahorkay:2015iga,Krasznahorkay:2017gwn,Krasznahorkay:2019lyl}
provides us with a candidate model that should be explored further,
with a possible impact on high-precision spectroscopy
and scattering experiments~\cite{Jentschura:2020zlr}.

Finally, in view of recent, mutually contradictory results
on the proton radius~\cite{Beyer:2017gug,Fleurbaey:2018fih,Bezginov:2019mdi,Xiong:2019umf},
the hope is expressed that future experiments
will be consistently evaluated.
This hope is being formulated taking into account a 
possible non-universality of electron-proton versus
muon-proton interactions~\cite{Camilleri:1969ag,Jentschura:2014ila,Jentschura:2014yla}. 

{\em Acknowledgments.---}Support from the National Science
Foundation (Grant PHY--1710856) is being gratefully acknowledged.

\ctitle{Summary EIC MC activities}
\byline{Markus Diefenthaler}

Markus Diefenthaler presented activities by the EICUG Software Working Group (SWG). The SWG has participated in the call for Expressions of Interest (EoI) by the EICUG and asked the community to present software needs for the EIC and how to meet this needs:
\begin{itemize}
    \item \textbf{Requirements} What software needs for EIC Software would you like to highlight now, in a few years, and for the completion of the EIC project?
    \item \textbf{Technologies \& Techniques} What software technologies and techniques should be considered for the EIC?
    \item What resources can your group contribute?
\end{itemize}

Based on the community input, the SWG submitted a Software EoI that will evolve towards a work plan, setting priorities for the next years and goals for the next decade. The SWG has reviewed Monte Carlo event generators (MCEGs) for ep and eA processes and discussed their requirements and developments for the science program at the EIC.

For the preparation of the TDR, the SWG will maintain a collection of MCEGs that are used in the EICUG and are validating them against existing DIS data. It will provide an online catalog of MCEGs where the community can select MCEGs and find documentation, validation plots, and examples on how to use the MCEG with existing EIC Software. The MCEGs will be fully integrated in the workflows for physics and detector simulations. Comparison to H1, ZEUS, and COMPASS measurements are being used to validate the MCEGs. For the MC-data comparison, the SWG is utilizing the RIVET tools as recommended by the worldwide MC community.

For the completion of the EIC project, the SWG will define with the community what part of the MCEG collection can be substituted and what part needs to be preserved for the EIC. It will make the necessary changes to the legacy MCEGs to ensure their future compatibility, e.g., adding HEPMC3 as output format and integrating them in the RIVET workflow. At the same time, the SWG will work with PARTONS and other initiatives to integrate new and upcoming MCEGs in the MCEG collection. The SWG will collaborate with the worldwide MCEG community on a global DIS tune based on our validation work. Merging of QED and QCD effects in MCEGs and correcting for these effects in the analysis will be essential. Please consider joining the SWG efforts and help with your expertise. 

The software development will build upon the infrastructure of the EICUG, including the GitHub organization and the GitHub website. Please consider storing (or mirroring) your source code and examples, e.g.\ on unfolding algorithms, in the EICUG GitHub organization (\url{https://github.com/eic}). 

\ctitle{Second-Order Radiative Corrections for ep Scattering}
\byline{Hubert Spiesberger }

First order radiative corrections for electron proton scattering  
have been known to be large since long. They are enhanced by 
large logarithms of the ratio of the momentum transfer $Q^2$ 
and the electron mass $m_e$, $\ln(Q^2/m_e^2)$, due to radiation 
of photons collinear with the incoming or scattered electron. 
In addition, a strong enhancement of radiative corrections is 
possible due to the fact that the energy carried away by a 
radiated photon leads to a shift of the momentum transfer to 
small values. This radiative tail is particularly important 
in inelastic scattering at large inelasticity $y$, reaching 
corrections in the order of 100~\%, but is also important 
for elastic scattering. 

One should therefore expect that the suppression due to 
additional powers of $\alpha/2\pi$ of corrections beyond the 
first order is counteracted by the presence of additional 
powers of a large logarithm. Kinematic cuts on the observed 
scattered electron or on the allowed photon energy can lead 
to an additional enhancement. Radiative corrections therefore 
depend strongly on the details of the experiment and should 
be studied with the help of Monte Carlo event generators which 
can simulate the experimental conditions.

\begin{figure}[b]
\begin{center}
\includegraphics[width=6cm]{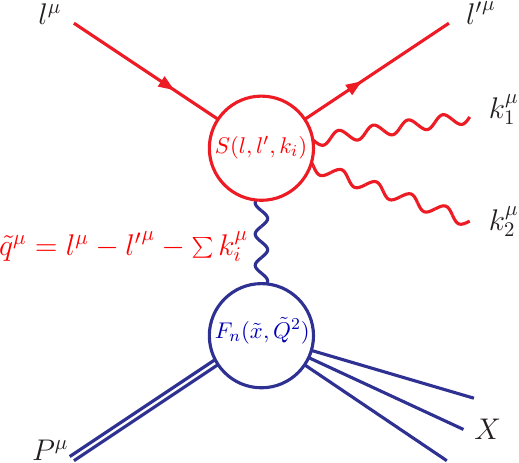}
\caption{
\label{Fig:src-radiator}
Leptonic radiative corrections to lepton-proton scattering 
can schematically be described as a convolution of a radiator 
function $S(l, l^\prime, k_i)$ due to photon emission 
from the lepton line with structure functions or form 
factors $F_n(x, Q^2)$ describing the hadron structure. 
The radiator function $S$ includes $\delta$-function terms 
from loop corrections.  
}
\end{center}
\end{figure}
 
Leptonic corrections are the dominating contribution to 
higher-order corrections. Since this contribution is 
gauge invariant by itself, it can be calculated separately. 
It is described by Feynman diagrams where photons are 
attached in all possible ways to the lepton line, see 
Fig.~\ref{Fig:src-radiator}. At second order one has to 
take into account two-loop diagrams, one-loop corrections 
to radiative scattering, and diagrams for the emission of 
two photons. Infrared divergences, present in each part 
separately, are cancelled when real and virtual corrections 
are combined. Details of the calculation can be found in 
Ref.~\cite{Bucoveanu:2018soy} (see \cite{Banerjee:2020rww} 
for an alternative approach).

In the limit of vanishing photon energy, i.e.\ in the 
soft-photon limit, higher-order corrections can be estimated 
by exponentiation. Also logarithmically enhanced contributions 
due to hard collinear photons can be calculated approximately. 
A convenient approach is based on the QED analog of parton 
splitting functions known from QCD, see for example  
\cite{Kripfganz:1990vm,Blumlein:2007kx}. These leading-log 
calculations are based on the assumption that radiation is 
strictly collinear, i.e.\ on the peaking approximation. 
Forthcoming high-precision measurements, however, require 
a calculation which goes beyond these approximate approaches.

\begin{figure}[b]
\begin{center}
\includegraphics[width=\linewidth]{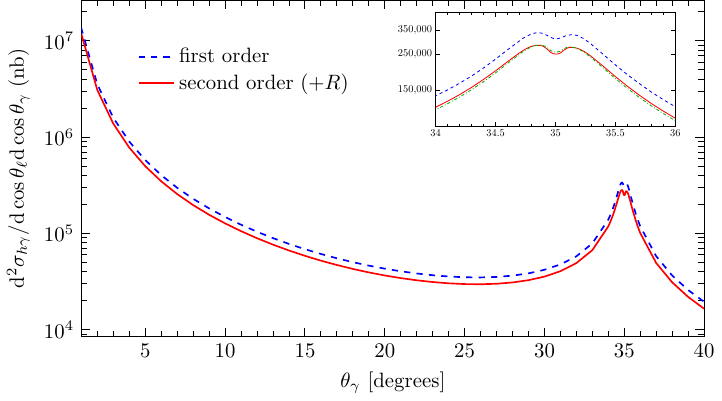}
\caption{
\label{Fig:src-radxsection}
Differential cross section for radiative ep scattering as 
a function of the photon emission angle at leading order 
(blue dashed curve) and including $O(\alpha)$ corrections 
(red full curve). The beam energy is $E = 155$~MeV and the 
electron scattering angle is fixed at $\theta_e = 35^\circ$. 
The radiated photon is allowed to have an energy between 
10~MeV and the kinematic maximum. 
}
\end{center}
\end{figure}

As an example we show in Fig.~\ref{Fig:src-radxsection} 
the cross section for radiative ep scattering, i.e.\ with 
an additional photon in the final state \cite{Bucoveanu:2018soy}. 
After integrating over the photon phase space, this contributes 
a correction of order $O(\alpha)$ to elastic ep scattering. We 
have chosen the kinematic variables at values which are relevant 
for the P2 experiment at MESA, Mainz, \cite{Becker:2018ggl} 
i.e.\ an electron beam energy of $E = 155$~MeV and a scattering 
angle of $35^\circ$. We have integrated over the photon energy 
imposing a lower cutoff $E_\gamma \geq 10$~MeV. The graph 
exhibits two prominent peaks, one at small angles where the 
photon is emitted into a direction close to the initial 
electron, and one at $35^\circ$ where the radiated photon is 
close to the scattered electron. Radiative scattering appears 
with a substantial tail to angles far away from the incoming 
or scattered electron. It is obvious that a peaking 
approximation, based on the assumption that radiation 
is strictly collinear, is not appropriate if one requires 
high-precision.

\begin{figure}[t]
\begin{center}
\includegraphics[width=0.8\linewidth]{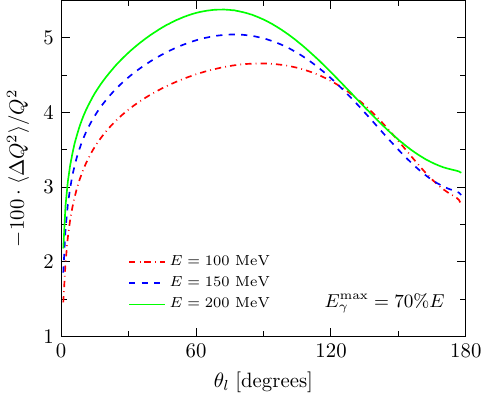}
\caption{
\label{Fig:src-avshift}
Average relative shift of the momentum transfer 
(in percent) for three values of the electron beam energy, 
assuming that a radiated photon can take away up to 70~\% 
of the beam energy. 
}
\end{center}
\end{figure}

\begin{figure}[t]
\begin{center}
\includegraphics[width=\linewidth]{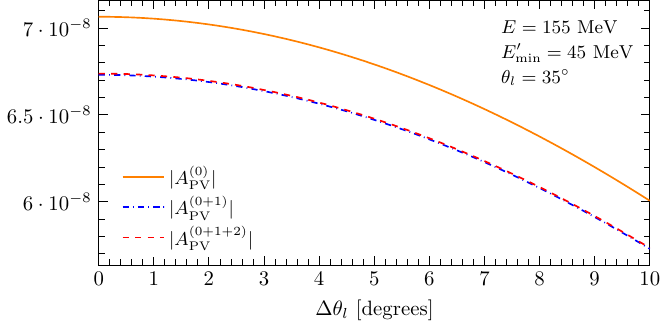}
\caption{
\label{Fig:src-APV} 
The parity-violating asymmetry for elastic ep scattering of 
polarized electrons at the Mainz P2 experiment. The large 
corrections are due to the shift of $Q^2$ by photon radiation. 
Second-order corrections are small in this case. 
}
\end{center}
\end{figure}

For an interpretation of ep cross section measurements 
in terms of structure functions or form factors, a precise 
knowledge of the momentum transfer is needed. If a radiated 
photon stays unobserved, the true momentum transfer to the 
proton can not be calculated from the scattering angle of 
the electron. This kinematic effect can amount to several 
percent, as shown in Fig.~\ref{Fig:src-avshift}. It is 
important for the extraction of the weak mixing angle 
from a measurement of the parity-violating helicity 
asymmetry, as planned at the P2 experiment at MESA in 
Mainz \cite{Becker:2018ggl}. Figure \ref{Fig:src-APV} shows 
the result of a calculation at leading order, as well as one 
including first-order and second-order corrections. In this 
case, one finds only small corrections at order $O(\alpha^2)$.  

While the theoretical treatment of leptonic QED radiative 
corrections is well established, uncertainties remain when 
hadronic effects come into play. A full calculation will have 
to include also corrections due to photon radiation from the 
hadron side, as well as lepton-hadron interference effects 
combined with two-photon exchange contributions. They introduce 
an additional model dependence since they require knowledge of 
the hadron structure beyond the elastic form factors. These 
hadronic corrections are expected at the level of fractions 
of a per mille, much smaller than the leptonic corrections, 
but a verification of these expectations is needed for future 
high-precision measurements.

\ctitle{Higher-order calculations of RC for ep and Moller scattering}
\byline{Stan Srednyak}
Recently, the world value of the charge radius of the proton, as measured in ep collisions,  was reconsidered \cite{xiong2019small}. This experiment was motivated by the previous measurement of this quantity in muonic hydrogen spectroscopic experiment \cite{pohl2010size}. Although at the moment there seem to be less disagreement between the two types of measurements \cite{xiong2019small} (after correcting the world value that corresponds to ep type measurement), this experiment lead physicists to reconsider theoretical background on which the interpretation of the data from these experiments is based. The conclusions are threefold. First, the bound state equation that was used for two loop computation \cite{eides2001theory,czarnecki2005calculation, karshenboim2005precision,pohl2013muonic} for spectrum must be reconsidered because there are no first principles that substantiate this equation beyond one loop order. Second, on the scattering side, there are two photon exchange processes whose contribution can only be estimated, rather than calculated precisely \cite{Afanasev:2017gsk,Ahmed:2020uso, Tomalak:2018jak}, because currently theory does not provide means to evaluate the matrix elements of non local operators. Third, both these computations are very challenging from mathematical point of view, and the field would benefit greatly from the development of mathematical tools to facilitate these computations. 

Extraction of the proton charge radius from spectroscopy experiments  is based \cite{pohl2013muonic,eides2001theory} on the proportionality of the Lamb shift to this radius. This calculation is based on the Bethe-Salpeter equation. There is no explicit closed form solution to this equation. An approximation scheme (NRQED) was developed \cite{caswell1986effective, czarnecki2005calculation, eides2001theory} for systematic extraction of terms of different order in $\alpha, log(\alpha), m/M$. It is the expressions obtained by this scheme that are used to extract $r_p$ from experimental data \cite{pohl2013muonic}. It is not the direct prediction of the Bethe-Salpeter equation that are used, but rather the results of the approximation scheme. There is little doubt that the terms in this scheme were extracted correctly, as this computation was checked by multiple groups (see \cite{eides2001theory, czarnecki2005calculation, karshenboim2005precision, pohl2013muonic} and refs therein). Therefore, it is not the calculation that has to be reconsidered, but the basic principles on which this computation is based. The reconsideration must be twofold. First, it is known \cite{Srednyak:2018kxn, Kanatchikov:2018uoy, kiefer1992functional} that it is very hard mathematical problem to formulate gauge invariance in theories with multi-local terms, such as bound state equations. These equations must be functional, directly formulated in the full Hilbert space of the theory that involves towers of virtual particles. Unfortunately, activity in this direction was rather limited in the recent years \cite{Srednyak:2018kxn,Kanatchikov:2018uoy,brodsky1998quantum}, perhaps because it is very difficult to solve resulting equations even approximately (we wish to remind the reader that one-loop computation of leading terms to Bethe-Salpeter was completed \cite{pachucki1994complete} only after forty years after the importance of this problem was recognized \cite{Karplus:1952zza}. This equation is a toy model for the true functional equations for bound states.) Therefore, to make progress in this field, more resources should be allocated to the development of mathematical methods for the solution of such equations. Such methods would include, for example, the theory of hypergeometric functions on symmetric spaces \cite{hotta2007d,takeuchi2010monodromy} (see more below).

Extraction of proton charge radius from ep scattering experiments must be accompanied by the evaluation and subtraction of radiative corrections to scattering. This problem is well defined except for the two-photon exchange process, which is currently model dependent. In PRad \cite{xiong2019small}, this correction was only estimated. Therefore, to advance the interpretation of the results of PRad I and the reduction of error bars on $r_p$ in this experiment, a new calculation of the radiative corrections is necessary. Moreover, it is desirable to advance these calculations to two loop order to match the accuracy on the spectroscopy side. This is a very difficult problem, as clearly recognized by the experts in the field \cite{Banerjee:2020tdt, DiVita:2019lpl, Frellesvig:2019uqt}.

The problem of evaluating two-photon exchange contribution is a fundamental one, drawing attention for long time (see the review \cite{Afanasev:2017gsk} and refs therein, and modern \cite{Ahmed:2020uso, Tomalak:2018jak}). However, all these papers aim either at estimation of this effect (for example, by using dispersive methods \cite{Blunden:2017nby} or model dependent vertices \cite{Tomalak:2018jak}) or extraction of parameters form e.g.\ charge asymmetry experiments \cite{Rachek:2015ymm}. The problem of evaluating this contribution in continuum QCD is largely considered intractable. It is related to the problem of finding the bound state equations for gauge theories and solving them at strong coupling. Although clearly inaccessible at this point, it draws attention to the development of powerful mathematical methods suitable for the solution of functional partial differential equations, and for systematic analysis of such equations. This latter research can be very fruitful in our opinion, as we detail later.

The evaluation of one loop corrections to ep scattering has been considered in many papers (see \cite{Akushevich:2019mbz} and refs. therein). Extending this success story beyond one loop probably will lead to a disappointment because the corresponding calculations are very difficult \cite{Banerjee:2020tdt, DiVita:2019lpl, Frellesvig:2019uqt}. The long standing problem here consists in the development of suitable mathematical tools for the evaluation of multiloop integrals. There are many groups working in this direction. Let us mention only a few of them. In \cite{adams2013two, bloch2015elliptic}, two-loop sunrise was investigated and it was shown that it leads to a new class of functions beyond the class of polylogarithms - so-called elliptic polylogarithms. There are deep connections with arithmetic algebraic geometry and the theory of motives \cite{bloch2015feynman}. General approach to multidimensional integrals was formulated long time ago in mathematical literature \cite{gelfand1990generalized}. There are some problems in adapting this approach to the physical situation, the main of them being the understanding the complicated geometry of singularity divisor of Feynman integrals. This approach has been pursued by the author \cite{Srednyak:2018hes}, and it is hoped that it can lead to concrete numerical schemes suitable e.g.\ at two-loop order. This approach lead to the representation of Feynman integrals in the form 
\vskip-5ex\begin{gather}
J_a=\sum L_i^{\alpha_{a,i}} g_{a,i}(q_p)
\label{eq:1}
\end{gather}
\vskip-1ex
where $L_i$ are Landau polynomials and $g_{a,i}(q_p)$ are analytic functions depending only on the coordinates on the locus $L_i=0$, for which a similar expansion holds. The coefficients in this expansion can have poles in $d-4$. Therefore, the function of all the kinematical invariants can be decomposed into a tower of functions depending on progressively less variables. It is still a difficult problem to evaluate the coefficients in this expansion. One of the approaches to this problem involves pull-back of holonomic D-modules and b-functions \cite{oaku1997algorithms}. The general paradigm of application of this approach consists in 1) finding expressions for the Landau polynomials 2) finding the set of master integrals 3) finding the exponents $\alpha_{a,i}$ for each landau polynomial and each master integral 4) stratification of the parameter space by multiple intersections of Landau polynomials 5) expanding iteratively master integrals and coefficient functions $g_{a,i}$ into the Gamma-series of the above type. Each of these steps is non trivial. For example, even the evaluation of Landau polynomials leads to expressions of very high degree, as in general it involves determinants of discriminantal complexes \cite{gelfand1994discriminants,dolotin609022introduction}. The next problem is the description of singularities of their intersections. This problem is not solved for two loop integrals (it probably leads to a set of interesting arithmetic varieties). Although it is possible to use the Landau polynomials as coordinates on the space of kinematic invariants, it may be advantageous to study different algebro-geometric models of this space, for example, resulting from various blow-ups of this space \cite{spivakovsky1983solution}. The problem of finding analytic expression for sections of flat $GL(n,C)$ bundles has also been considered from moduli point of view \cite{hitchin1987stable} (so-called Higgs bundles in the case of Riemann surfaces). Feynman diagrams correspond to very specific high dimensional representations of the fundamental group of the parameter space. Therefore, the question arises, how do these bundles correspond to points in the moduli space of general rank $r$ bundles on the parameter space. The approach to this problem is provided by isomonodromy approach. Unfortunately, the isomonodromy theory is well developed only in the case of 1d parameter space, while we need it for multiple dimensions (already in 1d case, it is much more complicated when the Riemann surface is different from the Riemann sphere). Therefore, a very bold extension of the existing methods must be accomplished in order to set this problem on rigorous mathematical basis. A typical result that one should strive for in this area may consist in classification of stratified spaces arising in on-shell 2-2 scattering, and classification of Pfaffian forms of regularly singular PDEs on such spaces. 

Recent developments \cite{bloch2015feynman} substantiate the exciting realization that the multiloop integrals are in fact number theoretic objects. Number theory enters through the geometry of the singularity locus. There are number fields associated to particular multiloop diagrams. These number fields are generated by coordinates of zero-dimensional strata in the stratification by intersections of Landau polynomials. What is the physical meaning of this phenomenon (if any ) remains to be explored. 

As we mentioned, the development of continuum QCD will entail development of techniques to solve functional partial differential equations \cite{Srednyak:2018kxn, Kanatchikov:2018uoy, brodsky1998quantum}. One of the simplest such equations is the following functional Laplacian
\vskip-4.5ex\begin{gather}
\int dx dy K(x,y)\frac{\delta^2} {\delta \phi(x) \delta \phi(y)} F[\phi]=0
\end{gather}
\vskip-1ex
for a given function $K(x,y)$. In the case of Bethe-Salpeter, this function involves $2-2$ kernel (and thus, at one loop, dilogarithms) but as a toy model one can take $K(x,y)=1/P(x,y)$ with a generic polynomial $P(x,y)$. Note that this equation is meaningful even in the case of 1d space. It is remarkable fact that the formal solution of this equation can be sought in the form of the series of the type eq.\ \ref{eq:1}. I.e., before we resum the series, individual terms are flat bundles on a complement of the zero locus of several polynomials, that are generated by the original form of $K(x,y)$. Flat bundles provide a kind of universal class of functions which solve a variety of problems in QFT. Resummation will call for a wide extension of this function class. It will be necessary to consider the class of functions that have logarithmic singularities along a locus given by zero set of an entire function of the kinematic invariants. This is work in progress by the author. 

The problem of continuum bound state formulation of QCD (and QED beyond one loop) leads to deep questions of the theory of functions of several complex variables. In particular, it is known that asymptotics at infinity (in x-space) of wave functions is exponential. This theory is well understood in 1d case \cite{balser1991multisummability}, but it is much less understood in multidimensional case, as needed for physics \cite{sabbah2012introduction, mochizuki2008wild, kedlaya2010good}. What is clear at the moment is that a multidimensional analogy of Stokes phenomenon is involved in understanding multiparticle states. The question lies in the description of the class of function in which to seek for solutions. A broad class of functions is provided by so-called resurgent functions \cite{mitschi2016divergent}. Remarkably, this same class shows up in the problems of quantum mechanics that involve chaotic phenomena \cite{ezra1991semiclassical}. As these are toy problems compared to the true bound state problems with radiative corrections, it is highly unlikely that it will be possible to avoid considering this broad function class.

\ctitle{Factorized approach to radiative corrections for inelastic lepton-hadron collisions}
\byline{Jianwei Qiu}

{\it Introduction---}\ 
We propose a new factorized approach to QED radiative corrections (RCs) in inclusive and semi-inclusive lepton-hadron deep-inelastic scattering (DIS), which treats QED and QCD radiation on an equal footing.
The method allows the systematic resummation of the logarithmically enhanced RCs into factorized lepton distribution and fragmentation (or jet) functions that are universal for all final states.
The new approach provides a uniform treatment of RCs for the extraction of parton distribution functions, transverse momentum dependent distributions, and other partonic correlation functions from lepton-hadron collision data~\cite{Liu:2020rvc}.

Instead of treating QED radiation as a correction to the Born process~\cite{Mo:1968cg, Bardin:1988by}, which becomes increasingly difficult beyond inclusive DIS~\cite{Ent:2001hm, Afanasev:2002ee, Akushevich:2019mbz}, we unify the QED and QCD contributions in a consistent factorization formalism.
The new approach allows all collinear (CO) sensitive and logarithmically-enhanced photon radiation to be systematically resummed into factorized, either CO or TMD, universal lepton distribution functions (LDFs) and lepton fragmentation (or jet) functions (LFFs), or as QED modifications to the evolution of nucleon PDFs or TMDs.
The remainder of the QED contributions are then insensitive to lepton mass $m_e \to 0$, and included in the infrared (IR) safe, perturbatively calculable hard parts, plus corrections suppressed by powers of $m_e$ over the hard scale of the collisions.
Calculating RCs is then replaced by determining the universal LDFs and LFFs, along with perturbative calculation of IR-safe QED corrections to the short-distance hard parts.
With both QCD and QED factorization, the new framework provides a consistent and perturbatively stable strategy to extract PDFs and TMDs from lepton-nucleon scattering.

{\it Inclusive lepton-nucleon DIS---}\ 
In the absence of photon radiation from leptons, and dropping lepton and hadron masses, the inclusive cross section for lepton-nucleon DIS $e (\ell) + N (P) \to e (\ell') + X$, is given by
\begin{eqnarray}
\label{e.indis0} 
E' \frac{\diff \sigma_{\mbox{\tiny{\rm DIS}}}}{\diff^3 \ell'}
&=& 
\frac{4 \alfa^2}{s \xb y^2 Q^2}
\Big[ \xb y^2 F_1 + (1-y) F_2 \Big],
\end{eqnarray}
where $\alfa$ is the electromagnetic fine structure constant,
and the kinematic variables are given by
    $s = (\ell + P)^2$,
    $Q^2 = -q^2 = -(\ell - \ell')^2 > 0$,
    $\xb = Q^2 / 2P\cdot q$, and
    \mbox{$y = P\cdot q / P\cdot\ell$}.
The structure functions $F_1$ and $F_2$ 
can be factorized in terms of PDFs with corrections suppressed by powers of $1/Q^2$~\cite{Collins:1989gx}.
The inclusive cross section \eqref{e.indis0} would then provide a clean probe of $F_1$ and $F_2$~\cite{Chekanov:2001qu, Abramowicz:2015mha, Benvenuti:1989rh, Adams:1996gu, Arneodo:1996qe, Airapetian:2011nu, Tvaskis:2010as}.

In the presence of photon radiation, however, $F_1$ and $F_2$ are not direct physical observables.
Without accounting for all photon radiation, the exchanged photon momentum $q$ cannot be determined exactly, which impacts the extraction of PDFs.
With the small $m_e$, photon radiation could be enhanced by $\ln(Q^2/m_e^2)$, leading to large, but not precisely determined, RCs that are sensitive to kinematic variables such as $\xb$ and $Q^2$.
The factorization approach proposed in Ref.~\cite{Liu:2020rvc}, while unable to ensure the extraction of $F_1$ and $F_2$, does nevertheless provide a perturbatively stable formalism for the reliable extraction of PDFs from inclusive DIS cross sections.

We consider the inclusive DIS cross section in Eq.~\eqref{e.indis0} for a lepton scattered with a large transverse component $\ell'_T \gg \Lambda_{\mbox{\tiny\rm QCD}}$ in the colliding lepton-nucleon frame.
Applying the factorization formalism for single-hadron production at large transverse momentum in hadronic collisions~\cite{Nayak:2005rt} to lepton-nucleon scattering, the factorized DIS cross section can be written,
\begin{eqnarray}
E'\frac{\diff \sigma_{\mbox{\tiny{\rm DIS}}}}{\diff^3 \ell'}
&=& \frac{1}{2s} \sum_{i,j,a} 
\int_{\zL}^1 \frac{d\zeta}{\zeta^2} 
\int_{\xL}^1 \frac{d\xi}{\xi}\, D_{e/j}(\zeta)\, f_{i/e}(\xi)
\notag\\
& & \hspace*{-0.7cm} \times
\int_{x_h}^1\frac{dx}{x} f_{a/N}(x)\,
\widehat{H}_{ia\to j}(\xi,\zeta,x;k')\
+\ \cdots
\label{e.fac}
\end{eqnarray}
where $i,j,a$ include all QED and QCD particles.
The lower limits $\zL, \xL$ and $x_h$ depend on external kinematics,
and the ellipsis represents $1/\ell^{\prime 2}_T$ corrections.
In Eq.~\eqref{e.fac}, the LDF $f_{i/e}(\xi)$ gives the probability to find lepton $i$ with light-cone momentum $\xi \ell^+$ in the incident lepton~$e$, while the LFF $D_{e/j}(\zeta)$ describes the emergence of the final lepton~$e$.
The LDF and LFF are defined in analogy with the quark PDF in the nucleon, $f_{a/N}(x)$, where $x=p^+/P^+$ is the nucleon momentum fraction carried by the quark, and quark to hadron fragmentation function~\cite{Collins:1981uw}, with the quark and gluon fields replaced by lepton and photon fields, and the hadron state by a lepton state.

In Eq.~\eqref{e.fac}, the $\widehat{H}_{ia \to j}$ is lepton-parton scattering cross section with all logarithmic CO
sensitivities along the direction of observed momenta, $\ell, \ell'$ and $P$, removed, and is therefore IR safe and insensitive to taking $m_e \to 0$.
The IR-safe $\widehat{H}_{ia \to j}$ can be perturbatively calculated by expanding the factorized formula \eqref{e.fac} order-by-order in powers of $\alpha$ and $\alpha_s$, with $\widehat{H}_{ia \to j}^{(m,n)}$ denoting the contribution at ${\cal O}(\alpha^m \alpha_s^n)$.
The factorized inclusive DIS cross section in Eq.~\eqref{e.fac} resums large photon radiation collinearly sensitive to the incident lepton into $f_{i/e}$, and radiation that is collinearly sensitive to the scattered lepton into $D_{e/j}$, and takes care of the photon radiation collinear to the colliding nucleon by adding QED corrections to the evolution kernels of the nucleon PDF $f_{a/N}$.

{\it QED contributions---}\
As for the factorized QCD case, the QED contribution to inclusive $eN$ scattering is organized into CO sensitive LDFs and LFFs, and IR-safe hard scattering functions.
Unlike the PDFs in QCD, however, which are nonperturbative, LDFs and LFFs are calculable perturbatively in QED. 
Focusing for brevity only on the ``valence'' lepton part [$i=j=e$ in Eq.~\eqref{e.fac}], at leading order (LO) in $\alpha$ we have $f^{(0)}_{e/e}(\xi) = \delta(\xi-1)$, and similarly the LFFs at LO are given by $D^{(0)}_{e/e}(\zeta) = \delta(\zeta - 1)$.
Explicit expressions for $f^{(1)}_{e/e}(\xi)$ and $D^{(1)}_{e/e}(\zeta)$ are given in Ref.~\cite{Liu:2020rvc}.
Expanding to lowest order [${\cal O}(\alpha^2 \alpha_s^0)$], we have
\begin{align}
&E'\frac{\diff \sigma_{\mbox{\tiny{\rm DIS}}}}{\diff^3 \ell'}
\approx \frac{2\alpha^2}{s} \sum_q 
\int_{\zL}^1 \frac{d\zeta}{\zeta^2} 
\int_{\xL}^1 \frac{d\xi}{\xi} D_{e/e}(\zeta)\, f_{e/e}(\xi)
\notag\\
& \hspace*{-0.3cm} \times
\int_{\xh}^1 \frac{dx}{x}\, e_q^2\, f_{q/N}(x)
\frac{x^2\zeta \big[ (\xi\zeta s)^2 + u^2\big]}
     {(\xi t)^2 (\xi\zeta s + u)}\,
\delta\big( x - \xh \big).
\label{e.fac0}
\end{align}
%
%
With the factorization formalism in Eq.~\eqref{e.fac}, one can systematically improve the ``RCs'' by calculating the IR-safe hard parts $\widehat{H}_{eq \to e}^{(m,n)}$ perturbatively for $m > 2$, and determining the 
universal, LDFs and LFFs.

{\it Numerical impact---}\
In general, the larger the momentum transfer $Q^2$, or more available phase space $s \gg Q^2$, the more important are the RCs.
In our factorization approach, the logarithmically enhanced RCs are included in the universal LDFs and LFFs, which shift the momenta of active leptons, and consequently impact the momentum transfer to the colliding nucleon.
To demonstrate the numerical impact of RCs, we show in \fref{sigmaratio} the ratio of the DIS cross sections
    $\sigma \equiv E' \diff\sigma_{\mbox{\tiny{\rm DIS}}}/\diff^3 \ell'$
without RCs (``$\sigma_{{\rm no\,}\mbox{\tiny {\rm RC}}}$'') and including RCs (``$\sigma_{\mbox{\tiny {\rm RC}}}$''), where 
    $\sigma_{{\rm no\,}\mbox{\tiny {\rm RC}}}$
is given by Eq.~(\ref{e.fac0}) with $f_{e/e} \to f_{e/e}^{(0)}$ and $D_{e/e} \to D_{e/e}^{(0)}$.
The cross section $\sigma_{\mbox{\tiny {\rm RC}}}$ 
is also given by Eq.~(\ref{e.fac0}) but evaluated with 
{\bf (i)}~NLO:
$f_{e/e} = f_{e/e}^{(0)} + f_{e/e}^{(1)}$ and 
$D_{e/e} = D_{e/e}^{(0)} + D_{e/e}^{(1)}$,
{\bf (ii)}~RES$_{\rm I}$:
resummed (or evolved) LDF and LFF with input distribution 
$f_{e/e} = f_{e/e}^{(0)}$ and $D_{e/e} = D_{e/e}^{(0)}$
at $\mu_0^2=m_e^2$, or
{\bf (iii)}~RES$_{\rm II}$:
resummed LDF and LFF with input 
$f_{e/e} = f_{e/e}^{(0)} + f_{e/e}^{(1)}$ and
$D_{e/e} = D_{e/e}^{(0)} + D_{e/e}^{(1)}$. 

\setlength{\abovecaptionskip}{0pt}
\setlength{\belowcaptionskip}{-10pt}
\begin{figure}
\includegraphics[width=0.5\textwidth]{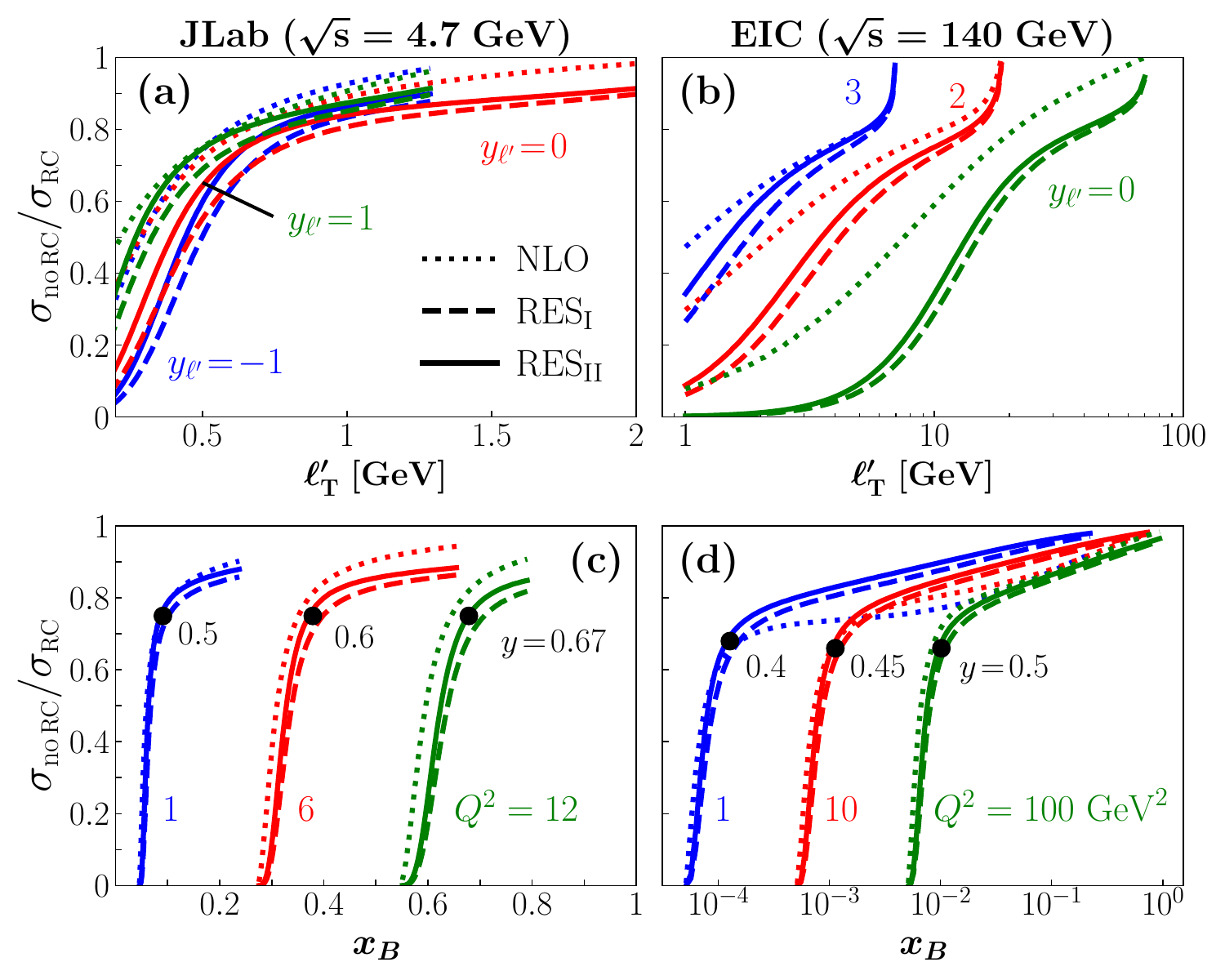} 
\caption{Ratios of inclusive $ep$ cross sections with no RCs to those including RCs computed according to the NLO (dotted lines), ${\rm RES_I}$ (dashed lines), and ${\rm RES_{II}}$ (solid lines) schemes, referred to in the text~\cite{Liu:2020rvc}. The values of $y$ at which the RCs are $\gtrsim 30\%$ are marked by the black filled circles.}
\label{fig:sigmaratio}
\end{figure}
\setlength{\abovecaptionskip}{10pt}
\setlength{\belowcaptionskip}{10pt}

The results in \fref{sigmaratio}, shown for kinematics typical of Jefferson Lab experiments and those planned for the EIC, illustrate some surprising and rather dramatic effects of RCs in certain regions of phase space.
Defining $y_{\ell'}$ to be the scattered lepton rapidity, the ratio $\sigma_{{\rm no\,}\mbox{\tiny {\rm RC}}}/\sigma_{\mbox{\tiny {\rm RC}}}$ versus $\ell'_T$ was shown for several fixed values of $y_{\ell'}$ typical for JLab [\fref{sigmaratio}(a)] and EIC [\fref{sigmaratio}(b)] kinematics.
The general trend of the effects is an increase in the magnitude of the correction at lower $\ell'_T$, consistent with more phase space available for photon radiation.
With the greater phase space available at the EIC, the RC effects can become quite significant at lower $\ell'_T$ values, even with $\ell'_T > 1$~GeV, and the differences between the RC prescriptions (NLO, $\rm RES_I$ or $\rm RES_{II}$) become more evident.
While the differences between the NLO and resummed results can be large, the  differences between the two resummed versions, $\rm RES_I$ and $\rm RES_{II}$, are much smaller.
This suggests that the resummation effects are important and the uncertainty in choosing the input distributions is mild, which provides greater stability in our factorization approach to calculate QED contribution to inelastic lepton-hadron scattering.

The impact of the RCs can be more directly visualized in terms of the traditional DIS variables $\xb$ and $Q^2$, and in \fref{sigmaratio}(c)--(d) the ratio $\sigma_{{\rm no\,}\mbox{\tiny {\rm RC}}}/\sigma_{\mbox{\tiny {\rm RC}}}$ is shown versus $\xb$ at fixed $Q^2$.
The most striking feature for both JLab and EIC kinematics is the dramatic change of slope in the $\xb$ dependence, from relatively shallow at higher $\xb$, where the RCs are $\lesssim 20\%$, to a steep fall-off with decreasing $\xb$.
For JLab kinematics this turnover occurs when $y \approx 0.5-0.6$, depending on the $Q^2$ value, but at even smaller values, $y \approx 0.4$, for EIC kinematics.
Our results in Fig.~\ref{fig:sigmaratio} clearly indicate that extreme care should be taken when extracting partonic information from inclusive DIS data at low $\xb$ and high $Q^2$, in view of the potentially large uncertainties in the QED RCs.

{\it Outlook---}\
Our unified factorization approach to QED and QCD dynamics has important implications for future analyses of hard scattering at the EIC.
Our observation that, without being able to account for all radiation, the photon-hadron frame is not well-defined and the standard $\xb$ and $Q^2$ does not fully control the momentum transfer to the colliding hadron, could impact the precision with which one can extract TMDs and explore the nucleon's 3-dimensional landscape in momentum space.  
On the other hand, our unified factorization approach allows the systematic resummation of the logarithmically enhanced RCs into factorized LDFs and LFFs that are universal for all final states, leaving the fixed order QED corrections completely IR-safe and stable as $m_e\to 0$. 
In contrast to previous treatments of RCs computed for different observables, the new paradigm allows a uniform treatment of QED RCs for the extraction of PDFs, TMDs and other partonic correlation functions from lepton-hadron collision data.

\ctitle{Radiative Corrections in MC for EIC}
\byline{Andrea Bressan}

\begin{figure}[b]
 \centering
  \includegraphics[width=\columnwidth]{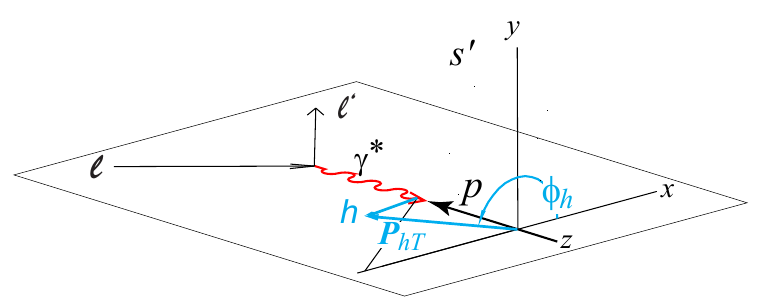}\hfill
  \caption{DIS scattering with hadron production in the Gamma Nucleon System}
\label{abfig1} 
\end{figure}

In the one-photon exchange approximation, and taking into account the transverse
momentum of the produced hadron $\vec{P}_{hT}$  the SIDIS cross-section
is differential in five independent
variables $\frac{d^5\sigma}{dx dy dz dP^2_{hT} d\phi_h}(\equiv d^5\sigma)$ \cite{Ji:2004wu,Bacchetta:2006tn};
in the unpolarized case, were we don't take into account neither the lepton nor the proton polarization 
we may write:
\begin{eqnarray}
d^5\sigma
&=& \frac{\alpha^2}{xyQ^2}\left[ ( 1-y ) + \frac{y^2}{2} \right] F_2 (x,
Q^2) \times \\
& & M_{UU}^h (x, Q^2,z,P^2_{hT}) \left\{ 1 + \right. \nonumber \\
& & \frac{2(2-y)\sqrt{1-y}}{1+(1-y)^2}
A_{UU}^{\cos \phi_h}(x,Q^2,z,P^2_{hT})\cos \phi_h + \nonumber \\
& & \left. \frac{2(1-y)}{1+(1-y)^2}
A_{UU}^{\cos 2\phi_h}(x,Q^2,z,P^2_{hT})\cos 2\phi_h \right\} \nonumber
\label{abeq1}
\end{eqnarray}
with the amplitudes of the $\cos \phi_h$ and $\cos 2 \phi_h$ ($A_{UU}^{\cos \phi_h}$ and $A_{UU}^{\cos 2\phi_h}$) 
modulations that are the ratio of semi-inclusive and inclusive form factors:   
\begin{equation}
A_{UU}^{\cos X\phi_h} = \frac{F_{UU}^{\cos X\phi_h}}{F_2}
\end{equation}
As can be seen from Fig.~\ref{abfig1}, we indicate the magnitude of the transverse momentum 
of the hadron $h$ with respect of the virtual photon $\gamma^*$ with $P_{hT}$, and with $\phi_h$ the azimuthal angle 
of the hadron-$\gamma^*$ plane with respect to the lepton $\ell$-$\ell'$ plane. Finally $z$ is the fraction
of the virtual photon energy taken by the hadron $h$. 

From the cross section in Eq.~\ref{abeq1} is therefore evident the importance of a correct evaluation 
of the direction of the virtual photon. All the azimuthal modulations (including the single spin 
asymmetries for the polarized case) are evaluated starting from this direction; and the same happens 
for the $P_{hT}$-dependent quantities, such as the $P_{hT}$-dependent multiplicities, indicated with 
$M_{UU}^h$ in Eq.~\ref{abeq1}. 

\begin{figure}[b]
 \centering
  \includegraphics[width=\columnwidth]{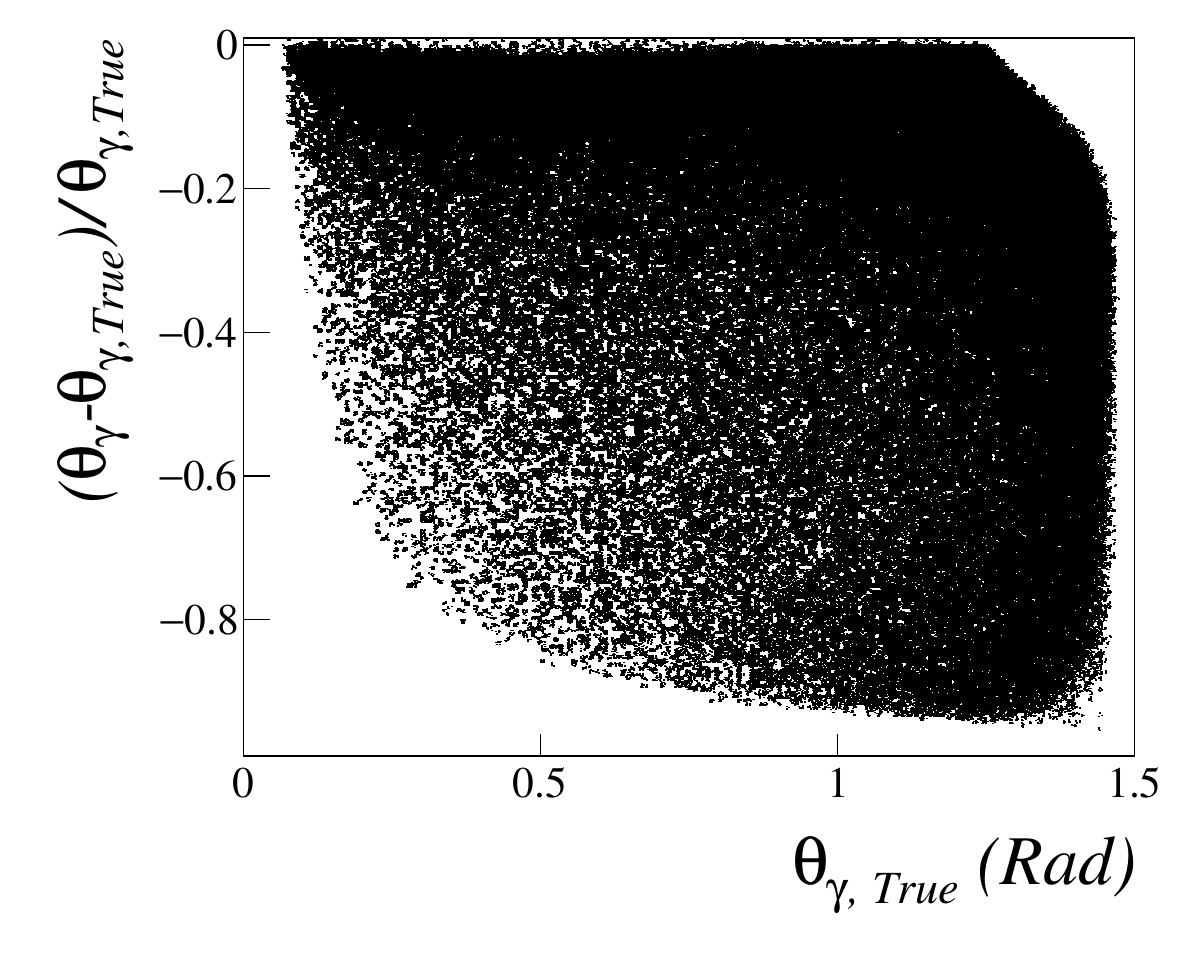}\hfill
  \includegraphics[width=\columnwidth]{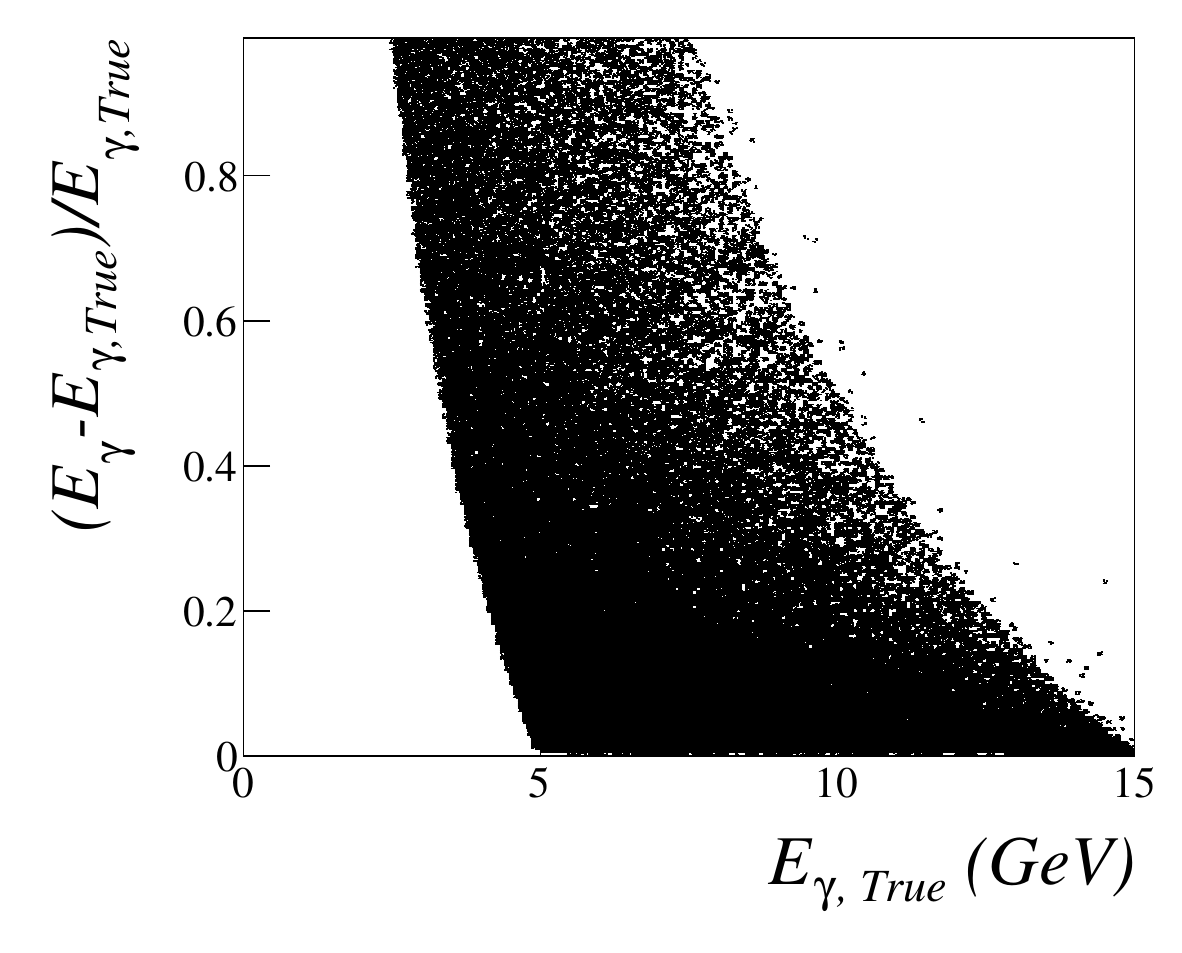}\hfill
 \caption{Relative difference between the polar $\theta_\gamma$ angle of the virtual photon with respect 
 to the incoming lepton beam in the Laboratory system protons as a function of $\theta_\gamma$ in the Lab 
 for the scattering of 5 GeV electrons on 50 GeV (left). Same for the virtual photon energy but in the GNS 
 (right).}
\label{abfig2} 
\end{figure}

Photon radiation from the lepton lines changes the DIS kinematics on the event-by-event basis. 
The true virtual photon direction and momentum are different than the one reconstructed from 
the leptons. This introduces false asymmetries in the azimuthal distribution of hadrons as 
calculated with respect to the reconstructed  hadron-$\gamma^*$ plane and smears the the kinematic 
distributions of $z$ and $P_{hT}$.
Due to the energy unbalance, the reconstructed energy of the $\gamma^*$ is always larger 
than the true one independently of if the photon was radiated by the incoming or by the outgoing 
lepton, in the same way as the polar angle of the virtual photon with respect of the incoming muon 
in the Laboratory System is always larger for the reconstructed $\gamma^*$ (as an example see in 
Fig.~\ref{abfig2} the results for scattering of 5 GeV electrons on 50 GeV).

The bias in the reconstructed virtual photon direction introduce quite sizable azimuthal modulations 
modulations, which impact on TMDs entering the unpolarized cross section. Since also the experimental 
apparatus may show some modulation in the acceptance, it is important to fold together a Monte Carlo 
encoding radiative effects in full simulations to calculate this experimental acceptance since these
modulations will mix together in the real data.

\ctitle{Opportunities at EIC with Both Lepton Signs}
\byline{Richard Milner}

Radiative corrections have been calculated predominantly in a single photon approximation with simple estimation of multi-photon contributions until the beginning of this century, when a large discrepancy was observed between cross section and recoil polarization determinations in the GeV region of the ratio of the proton's elastic form factors.  The widely accepted explanation is that hard two-photon effects have been neglected and affect the cross section determination.  Measurements of the positron to electron cross section ratio have been carried out DESY, Jefferson Lab and VEPP-2 up to Q$^2$ of about 2.4 (GeV/c)$^2$ but the results are inconclusive - see Axel's contribution here.  Further measurements to higher Q$^2$, where the contribution to the ratio should be larger, are under consideration at DESY by the TPEX collaboration and at Jefferson Lab, if a positron source can be realized. 

Given that multi-photon effects are accepted to be essential in understanding GeV elastic electron-proton scattering, the fundamental process in hadron structure, it is certainly possible that these will also be significant for the new processes planned for study at EIC. 

In addition, there are strong arguments to motivate the capability to have EIC positron beams based on compelling, open questions in hadron structure.  Here, the charged current (CC) interaction, shown in Fig.~\ref{fig:CC}, mediated by exchange of W$^+$ (W$^-$), becomes accessible by hard scattering of a positron (electron) from the hadronic target where the final-state lepton is a (undetectable) neutrino. Highlights include:

\begin{figure}[htpb]
    \centering
    \includegraphics[width=0.5\columnwidth]{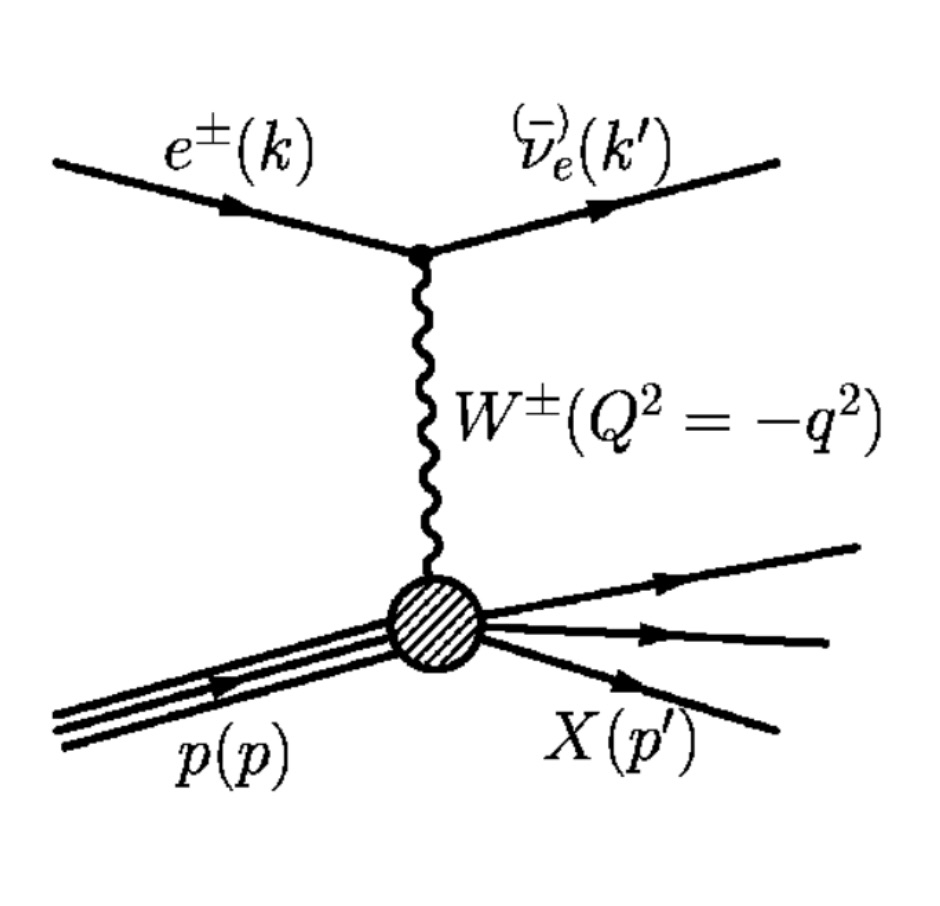}
    \caption{Feynman diagram for the charge current electroweak interaction.}
    \label{fig:CC}
\end{figure}

\begin{itemize}
    \item Flavor separation \\
    Using the CC electroweak interaction, it is possible to probe the $d/u$ ratio at large $x$, the light quark sea asymmetry and the strange/antistrange quark parton distribution functions.  
    \item Electroweak structure functions\\
    Using spin-dependent CC scattering, new structure functions, arising from W$^\pm$ exchange, can be accessed that involve {\it charmed}, {\it strange} as well as {\it up} and {\it down} quarks. Further, as with the Bjorken Sum Rule for {\it up} and {\it down} quarks, there are also fundamental sum rules for these new structure functions.  Experimental measurement of these sum rules can test our understanding of QCD.
    \item Deeply Virtual Compton Scattering(DVCS)\\
    The DVCS process is the leading deeply virtual exclusive process for pursuing new insights into hadron structure with EIC. It will allow access to new Generalized Parton Distributions and has the potential to give insight into the origin of nucleon spin, if Ji's sum rule can be determined. In particular, measurement of the beam charge asymmetry for DVCS (pioneered by the HERMES experiment) uniquely allows access to the real part of the Bethe-Heitler-DVCS interference term. 
\end{itemize}

In summary, as changing the lepton sign is key to experimentally accessing multi-photon contributions, measurement of the CC electroweak process and the full exploration of the DVCS process, it seems prudent to build the capability to have positron as well as electron beams into the next-generation, billion \$-class electron scattering facility.

\clearpage
\bibliography{whitepaper.bib}

\end{document}